\documentclass[twocolumn,floatfix,showpacs,floats,aps,pre,showpacs,amsfonts,amssymb]{revtex4}
\usepackage{dcolumn}
\usepackage{bm}
\usepackage{hyphenat}
\usepackage{graphicx}
\usepackage{psfrag}
\usepackage{amssymb}
\usepackage{algpseudocode}
\usepackage[latin1]{inputenc}
\newlength{\sepmod}
\def\eqref#1{(\ref{#1})}

\def\d{\mathrm{d}}
\def\sign{\mathrm{sign}\,}

\begin{document}
\title{Genetic embedded matching approach to ground states in continuous-spin systems}

\author{Martin Weigel}
\altaffiliation{Present address: Institut f\"ur Physik,
  Johannes-Gutenberg-Universit\"at Mainz, Staudinger Weg 7, 55099 Mainz, Germany}
\email{weigel@uni-mainz.de}
\affiliation{Department of Mathematics and the Maxwell Institute for Mathematical
  Sciences, Heriot-Watt University, Edinburgh, EH14~4AS, UK}
\affiliation{Department of Physics and Astronomy, University of Waterloo, Waterloo, Ontario,
  N2L~3G1, Canada}
\date{\today}

\begin{abstract} 

  Due to an extremely rugged structure of the free energy landscape, the
  determination of spin-glass ground states is among the hardest known optimization
  problems, found to be ${\cal NP}$-hard in the most general case. Owing to the
  specific structure of local (free) energy minima, general-purpose optimization
  strategies perform relatively poorly on these problems, and a number of specially
  tailored optimization techniques have been developed in particular for the Ising
  spin glass and similar discrete systems. Here, an efficient optimization heuristic
  for the much less discussed case of {\em continuous\/} spins is introduced, based on
  the combination of an embedding of Ising spins into the continuous rotators and an
  appropriate variant of a genetic algorithm. Statistical techniques for insuring
  high reliability in finding (numerically) exact ground states are discussed, and
  the method is benchmarked against the simulated annealing approach.

\end{abstract}

\pacs{75.50.Lk, 02.60.Pn, 75.10.Hk}

\maketitle

\section{Introduction}

Complex (free) energy landscapes featuring a multitude of local minima separated by
energy barriers are common in problems of statistical mechanics, chemical and
biophysics consequently often subsumed under the label of ``complex systems'', be it
biopolymers, structural or spin glasses \cite{wales:03}. The consequences of this
deviation from the classical textbook situation of a potential energy with at most a
handful of metastable states are dramatic for the static and dynamic behavior of the
affected systems in nature, including for instance ``memory'' and ``rejuvenation''
effects in spin glasses \cite{binder:86a}, but no less pronounced for the theoretical
investigation of models of such situations with computational simulation or
optimization methods: here, model systems get trapped in local minima for
exponentially long times, preventing an equilibration in finite-temperature
simulations \cite{binder:86a} or lead to a vastly increased effort needed for an
optimization procedure to yield ground states with finite probability
\cite{hartmann:book}. Clearly, the presence of many local minima alone is not
sufficient to pose serious problems for any optimization method more elaborate than a
strictly downhill, local search. Rather, it is the organization of minima and
interjacent barriers that is the cause for the trapping phenomena, and distinguishes
the milder from the more severe cases \cite{wales:03}. While, for instance, many
typical biopolymers exhibit landscapes with moderate barriers separating minima of
substantially different energies, with a ``funneling'' towards a unique global
minimum \cite{plotkin:02}, disordered and frustrated magnetic systems are rather
characterized by many (quasi-) degenerate minima close to the ground state(s)
separated by large barriers, leading to much more severe effects of metastability and
slow relaxation \cite{huse:86,garstecki:99}.

Independent of this connection between the structure of the energy landscape and the
real or artificial (Monte Carlo or optimization) dynamics of a frustrated system, the
problem of finding ground states or, alternatively, partition functions, of the
corresponding models has been considered as a problem in the field of computational
complexity \cite{mertens:02}. In computer science, traditionally mostly the ``worst
case'' complexities of algorithms have been considered, i.e., the asymptotic scaling
of the run time $T(N)$ with the problem size $N$ for the ``hardest'' set of input
data within the class of allowed inputs, $T_\mathrm{max}(N)$ \cite{papadimitriou:94}.
Quite generally, problems with an asymptotically polynomial form of
$T_\mathrm{max}(N)$ are considered tractable, whereas an exponential divergence for
the best known algorithm is associated with intractability.  Paradigmatic results
have been found for decision problems with ``yes'' or ``no'' answers, for which a
powerful classification scheme has been established: problems with a known polynomial
algorithm are grouped in ${\cal P}$, whereas a more general class of problems for
which the {\em correctness\/} of a solution can be checked in polynomial time is
denoted ${\cal NP\/}$. The potential hardness of the latter must then exclusively
result from the exponential growth of the search space, such that a theoretical
computer capable of infinite parallelism can solve such problems in polynomial time
\cite{mertens:02}.  The hardest ${\cal NP}$ problems, namely those whose polynomial
solution would imply polynomial complexity for all other ${\cal NP}$ problems, are
termed ${\cal NP}$ {\em complete\/}, which includes most of the well-known hard
problems such as the traveling salesman problem or the satisfiability problem. While
it is possible that all such problems might have polynomial-time solutions, i.e.,
${\cal P} = {\cal NP}$, this is now considered to be extremely unlikely, and ${\cal
  NP}$ problems almost certainly require an exponential computational effort
\cite{papadimitriou:94}. This classification extends to optimization (instead of
decision) problems where, specifically, those where the problem of deciding about the
existence of a solution better than a given bound in the cost function is ${\cal NP}$
complete, are termed {\em ${\cal NP}$ hard\/}.

For the Ising spin glass with Hamiltonian \cite{edwards:75a}
\begin{equation}
  \label{eq:EA_model}
  {\cal H} = -\sum_{\langle ij\rangle}J_{ij}\,\bm{S}_i\cdot\bm{S}_j,
\end{equation}
the problems of computing ground states or the partition function are known to be
${\cal NP}$ hard in space dimensions $d \ge 3$ or for two-dimensional (2D) systems in
a magnetic field \cite{barahona:82,bachas:84}. The zero-field 2D problem, on the
other hand, is tractable in polynomial time \cite{bieche:80a,saul:93,galluccio:00}.
In particular, ground states on planar graphs can be found by means of the mapping to
a minimum-weight perfect matching problem, as discussed below in
Sec.~\ref{sec:matching}.  While thus, generically, spin-glass ground-state problems
are hard, one has to keep in mind that this classification concerns the worst-case
behavior among all possible realizations of couplings $J_{ij}$ of the chosen
distribution (e.g., bimodal or Gaussian), whereas it is, for instance, simple to
specify the ground state of the purely ferromagnetic system with $J_{ij} = J_0 > 0$,
which also belongs to the allowed $J_{ij}$ realizations.  Hence, relevant for actual
computations is also the {\em average\/} complexity, depending on a chosen
probability distribution $P(\{J_{ij}\})$ of couplings. Within the spin-glass phase,
however, also this mean complexity is exponential for known exact approaches to the
problem in $d > 2$ \cite{liers:03}.  Correspondingly, heuristic optimization
techniques for finding low-lying or ground states are called for. These might include
generic approaches, such as simulated annealing \cite{kirkpatrick:84}, multicanonical
\cite{berg:94} or parallel tempering \cite{hukushima:96a} Monte Carlo simulations,
but also a number of methods specifically tailored to the problem
\cite{hartmann:96a,pal:96a,marinari:00,houdayer:01a,boettcher:01}, the latter
generally showing the best performance \cite{moreno:03,hartmann:book}. Insofar as
these methods make use of some type of relaxational (quasi-) dynamics, they to some
extent also suffer from the slowness of relaxation entailed by the structure of
free-energy minima and separating barriers. It should be pointed out, however, that
such slow dynamics is not equivalent to hardness in the classifications of
computational complexity \cite{grest:86}. Instead, slower than power-law relaxation
of local dynamics also occurs in computationally polynomial systems \cite{huse:86},
such as the Ising spin-glass model in two dimensions \cite{bieche:80a}. The
stochastic nature of most of these approaches requires a different description of
their time complexity or efficiency: since such methods do not guarantee to yield
ground states, one should now rather ask for the worst case or mean computational
effort to end up in a ground state with an {\em a priori\/} prescribed success
probability $p_s$ (for $p_s = 0.95$, say), or for the distribution (over disorder
realizations) of such minimal running times at fixed $p_s$. A framework for such
considerations will be developed below in Sec.~\ref{sec:performance}. Ground-state
searches for spin-glass systems are additionally complicated by an extraordinarily
broad distribution of ``hardness'' over disorder samples, which draws into question
the treatment of all samples with constant computational effort. In this context, it
is discussed below in how far properties of individual disorder samples can serve as
hardness indicators and hence an automatic effort adaptation can be achieved.

Ising spin-glass ground states have been considered with the aim to understand the
nature of the low-temperature phase while avoiding the equilibration problems of
finite-temperature simulations. Ground-state computations for systems with different
boundary conditions or with some fixed spins allow for the direct investigation of
domain-wall and droplet defects, whose properties should reveal in how far
finite-dimensional spin glasses are correctly described by mean-field theory (see
Ref.~\cite{kawashima:03a} for a review of recent developments). The polynomially
tractable 2D case, in particular, has provided a fruitful playground for testing
theoretical pictures of the spin-glass phase, and remains a topic of active research
to date \cite{amoruso:06a,fisch:06b,aromsawa:07,hartmann:07}. In terms of spin-glass
phases realized experimentally, in particular in the multitude of systems with
frustrating lattice structures that have come into focus more recently
\cite{greedan:01}, systems with {\em continuous\/} spins are probably more common
than the extremely anisotropic Ising case. In computing ground states for such
systems, modeled, say, by the Edwards-Anderson Hamiltonian (\ref{eq:EA_model}) with
continuous O($n$) spins $\bm{S}_i$, one leaves the relatively well-understood field
of {\em combinatorial\/} optimization. To my understanding, nothing is known about
the (suitably generalized) computational complexity of this problem. It is easily
seen \cite{hartmann:book,papadimitriou:94}, however, that already the $q$-state Potts
spin glass corresponds to a multi-terminal flow problem known to be ${\cal NP}$ hard
even in two dimensions for $q = 3$ \cite{ferrari:95}. Nothing would seem to indicate
that the {\em XY\/} case of continuous planar spins, or the Heisenberg model of
O($3$) rotators could be easier computationally than the discrete Potts
approximation.  With the exception of a study of the {\em XY\/} spin glass in the
Coulomb gas representation \cite{kosterlitz:99a}, all studies of low-lying metastable
states in O($n$) spin glasses (with $n>1$) have relied on variants of a simple
spin-quench technique corresponding to a $T=0$ Monte Carlo simulation with local
updates \cite{walker:80,morris:86a,kawamura:91,maucourt:98a} (apart from studies of
the computationally simpler case of the $n\rightarrow\infty$ spherical spin glass
\cite{lee:05a}). This spin quench follows from noting that a necessary condition for
metastability is that each spin be parallel to its local molecular field,
\begin{equation}
  \label{eq:internal_field}
  \bm{S}_i \parallel \bm{h}_i = \sum_j J_{ij}\bm{S}_j,
\end{equation}
leading to the prescription of an iterative alignment of single spins $\bm{S}_i$
parallel to $\bm{h}_i$. In contrast to the investigations of the Ising spin glass,
none of these approaches have allowed to find numerically exact ground states with a
reasonably high probability, such that, instead, effectively the properties of some
set of metastable states with unclear relation to the ground-state behavior have been
found and investigated. To improve on this, it is proposed here to combine exact
ground-state computations of Ising variables embedded into the continuous spins with
a specially tailored genetic algorithm exploiting the locally rigid cluster structure
of metastable spin configurations \cite{weigel:05f}. This results in an efficient
approach for ground-state computations of continuous-spin systems on planar graphs,
which is tested and assessed here for the case of the bimodal {\em XY\/} spin glass
on the square lattice, where numerically exact ground states can be found with high
reliability for systems up to about $30\times 30$ spins using currently available
computational resources. Implications of these results for the nature of the
low-temperature phase of this model have been discussed elsewhere
\cite{weigel:05f,weigel:06c}.

The rest of the paper is organized as follows. Section \ref{sec:matching} is devoted
to a description of the embedded matching technique for continuous spins, while in
Sec.~\ref{sec:genetic} the combination of this approach with a genetic algorithm with
cluster exchange is discussed. In Sec.~\ref{sec:performance}, the performance of this
approach for the 2D {\em XY\/} spin glass is investigated in terms of a detailed
statistical analysis, focusing on the large variations between disorder replica and
offering a standardized approach of ``quality assurance'' for stochastic optimization
algorithms. An exhaustive benchmarking of the new approach against general-purpose
techniques is not feasible. At least, however, results comparing to the simple
spin-quench used before and a more elaborate simulated annealing approach are
presented in some detail. Finally, Sec.~\ref{sec:concl} contains my conclusions.

\section{Embedded matching\label{sec:matching}}

For attacking the ground-state problem of continuous spin glasses in two dimensions,
inspiration is taken from the polynomial solution of the Ising problem, which is
hence described first, and then adapted to continuous spins via an embedding of Ising
variables.

\subsection{Ising ground states as perfect matchings}

\begin{figure}[tb]
  \centering
  \vspace*{-0.3cm}
  \includegraphics[clip=true,keepaspectratio=true,width=7cm]{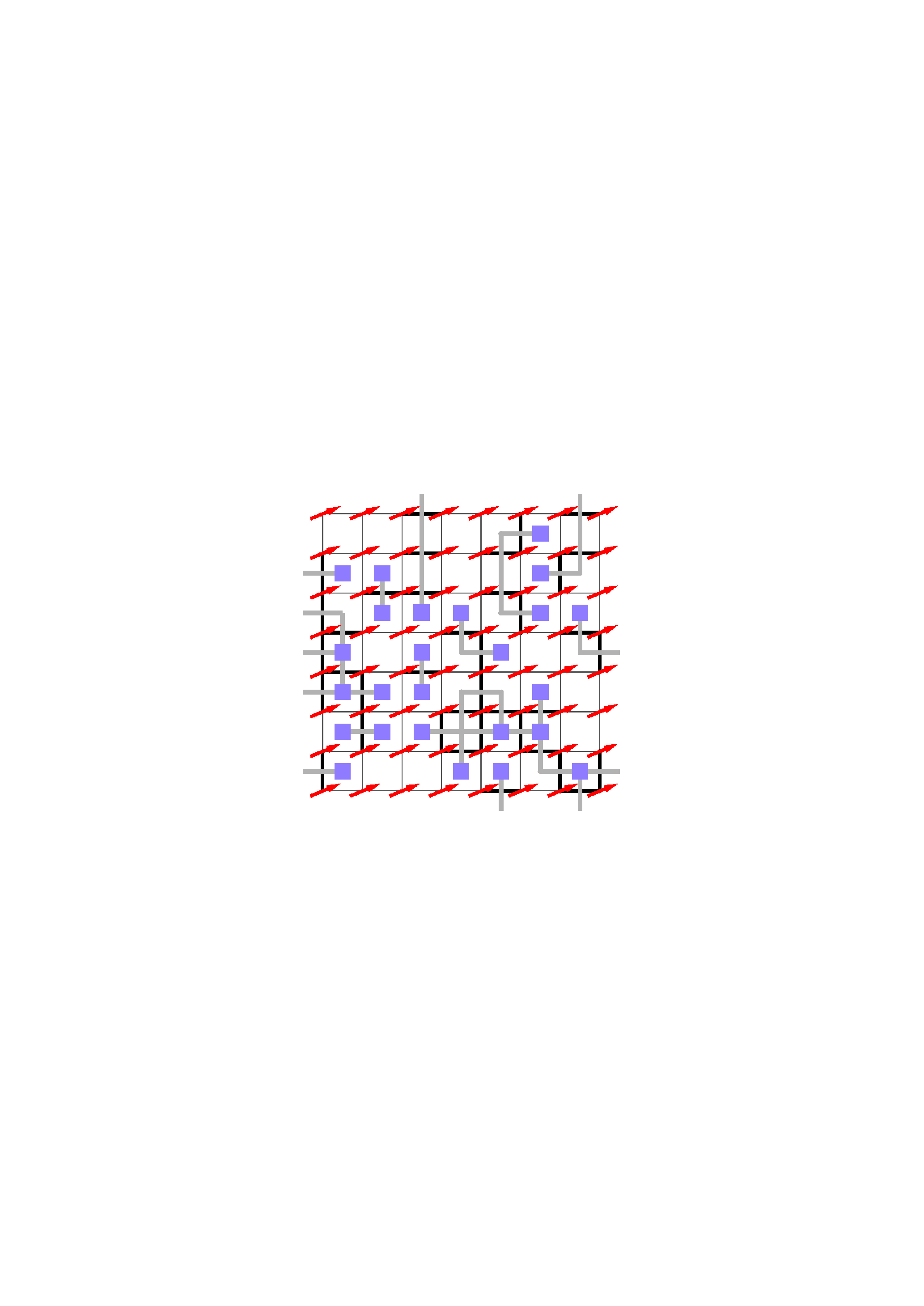}\\[-0.6cm]
  \includegraphics[clip=true,keepaspectratio=true,width=7cm]{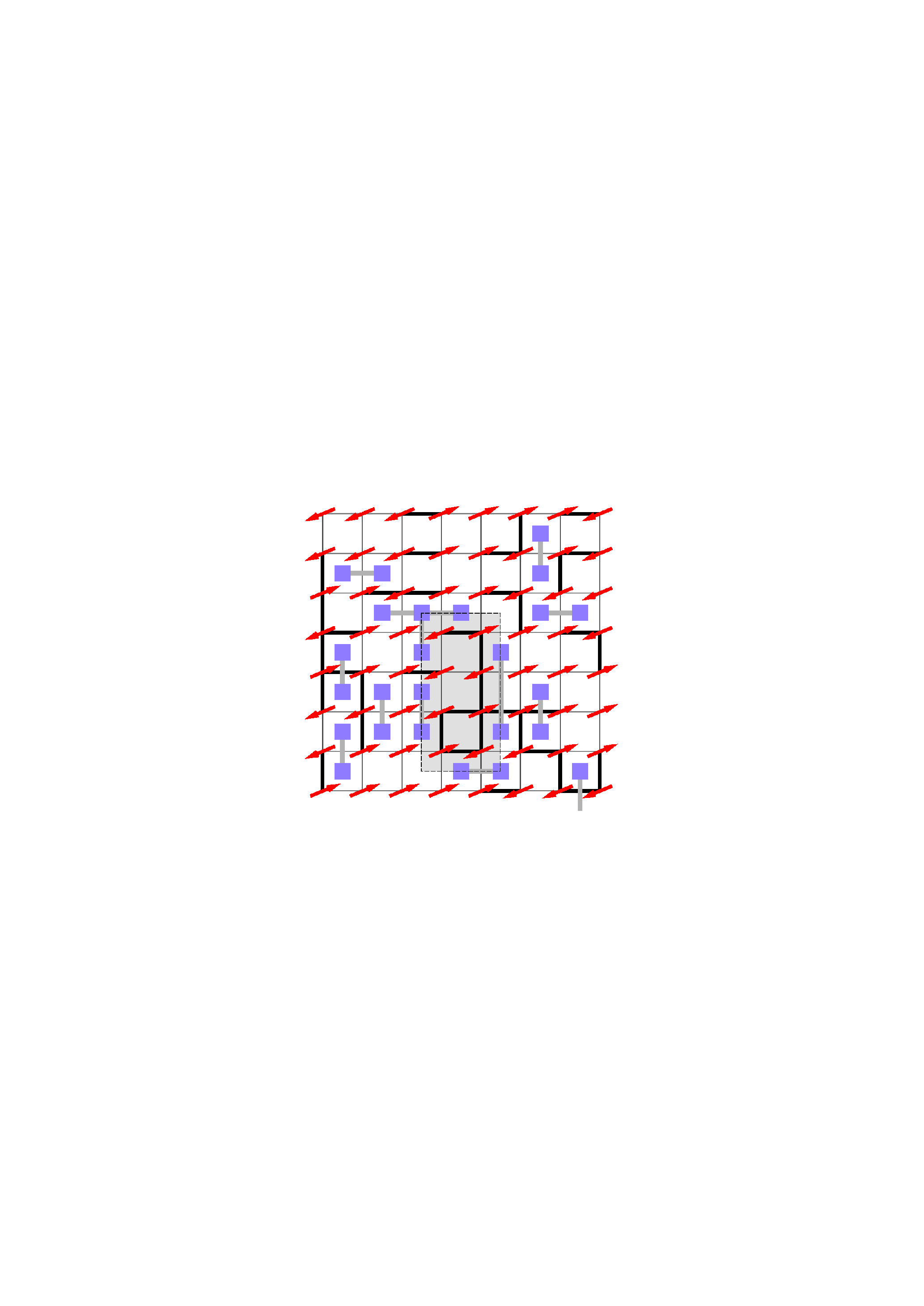}
  \vspace*{-0.3cm}
  \caption
  {(Color online) Transformation of the Ising ground-state calculation on the square
    lattice to a matching problem. Upper panel: Frustrated plaquettes (marked by
    small squares) have an odd number of antiferromagnetic (bold) bonds. The set of
    broken bonds forms a collection of lines on the dual lattice (shaded, gray
    lines). Lower panel: a ground state of the system is given by a minimum-weight
    perfect matching of frustrated plaquettes. The dashed line indicates an
    alternating cycle along which an exchange of matched and unmatched edges yields
    another perfect matching.}
  \label{fig:match}
\end{figure}

The polynomial complexity of the 2D Ising spin glass allows for the formulation of
efficient algorithms for finding ground states and computing the partition function.
A number of different techniques has been established for the calculation of the
latter \cite{barahona:82,saul:93,galluccio:00}, mostly relying on the computation of
Pfaffians, but these will not be needed here. Computations of ground states rest on
the concept of frustrated loops introduced by Toulouse \cite{toulouse:77a}: in the
presence of couplings $J_{ij}$ of either sign, for each closed curve along lattice
links touching an {\em odd\/} number of antiferromagnetic bonds ($J_{ij} < 0$), one
cannot find a spin configuration satisfying all pair interactions, i.e., $J_{ij}S_i
S_j < 0$ for at least one (``broken'') edge. Hence, the presence of such loops is
responsible for the excess of the ground-state energy of the spin glass above the
unfrustrated value $E_\mathrm{FM} = - \sum_{\langle ij\rangle} |J_{ij}|$.  Due to the
contractibility of all loops on a {\em planar\/} graph, in this case it suffices to
concentrate on the frustration of the {\em plaquettes\/}, i.e., the elementary faces
of the lattice \cite{bieche:80a}. This is illustrated by the marking of frustrated
plaquettes for the square lattice in Fig.~\ref{fig:match}. For each plaquette
$\square_n$, define the frustration function \cite{toulouse:77a}
\begin{equation}
  \label{eq:frustration}
  \Phi_{\square_n} = \prod_{(i,j)\in\square_n} \sign J_{ij} = \pm 1,
\end{equation}
such that $\Phi_{\square_n} = -1$ if and only if $\square_n$ is frustrated. By this
definition, in a configuration of the Ising spins each frustrated plaquette must have
an odd number ($1$ or $3$ for the square lattice) of broken bonds, whereas an
unfrustrated plaquette is surrounded by an even number of broken bonds ($0$, $2$ or
$4$ for the square lattice). Bonds drawn dual to the broken bonds then connect to
form {\em energy strings\/} starting and terminating in frustrated plaquettes, cf.\
the sketch in the upper panel of Fig.~\ref{fig:match}.  The excess energy is
\begin{equation}
  \label{eq:string_energy}
  \frac{1}{2}(E-E_\mathrm{FM}) = W_\mathrm{string} = \sum_\mathrm{strings}|J_{ij}|,
\end{equation}
and a ground state corresponds to a collection of strings of minimum weight
$W_\mathrm{string}$. Since, on a planar graph, closed loops of dual bonds can be
contracted away, they cannot occur in a ground state, which hence corresponds to a
{\em minimum-weight perfect matching\/} of the frustrated plaquettes. This is
illustrated in the lower panel of Fig.~\ref{fig:match}. The planarity of the lattice
ensures that each such matching corresponds to a valid spin configuration
\cite{bieche:80a}.

Following the above discussion, this matching problem is defined on the complete
graph ${\cal G} = {\cal F} \times {\cal F}$ on the set ${\cal F}$ of frustrated
plaquettes. Each of the $|{\cal E}| = |{\cal F}||{\cal F}-1|$ edges $e_{mn} = (f_m,
f_n)$ carries a weight
\begin{equation}
  \label{eq:dijsktra_weight}
  W(e_{mn}) = \min_{\gamma_{mn}} \sum_{(i,j)\in\gamma_{mn}} |J_{ij}|,
\end{equation}
corresponding to the minimum weight of all paths $\gamma_{mn}$ on the (original)
lattice connecting the plaquettes $f_m$ and $f_n$. Hence an auxiliary minimum-cost
path problem needs to be solved as an input to the matching calculation. This is most
efficiently achieved by an appropriate implementation of Dijkstra's algorithm with
O($|{\cal E}|\ln|{\cal E}|$) complexity, or, for the case of a bimodal P($J_{ij}$)
where $|J_{ij}| = J_0$ for all edges, by a simple breadth-first search
\cite{gibbons:book}.  Since there is an even number of frustrated plaquettes
\footnote{The product $\prod_n \Phi_{\square_n}$ over all plaquettes of the lattice
  is $+1$ for an even and $-1$ for an odd number of frustrated plaquettes. On the
  other hand, $\prod_n \Phi_{\square_n} = \prod_{\langle ij\rangle} (\sign J_{ij})^2
  = +1$, since each bond occurs twice in the product when taking into account
  external plaquettes for open boundaries.}, a perfect matching on ${\cal G}$ can
always be found. A polynomial algorithm for the matching problem on general graphs
has been proposed by Edmonds \cite{edmonds:65a}. It proceeds by successively
identifying augmenting paths in the matching graph, i.e., cycles of alternating
matched and unmatched edges such that an exchange matched $\leftrightarrow$ unmatched
decreases the overall weight. This is illustrated by a cycle in the original lattice
in the lower panel of Fig.~\ref{fig:match}. The complexity of the original
implementation is O($|{\cal F}|^2 |{\cal E}|$). The present implementation is based
on the ``Blossom IV'' matching algorithm of Cook and Rohe incorporating many
improvements developed in the combinatorial optimization literature after Edmonds'
original proposal \cite{cook:99a}.

Given a solution to the matching problem, a corresponding spin configuration is found
by arbitrarily choosing the orientation of one spin and successively implementing the
satisfaction constraints expressed by the perfect matching via selecting spin
orientations in a breadth-first search emanating from the chosen starting point. Note
that, depending on the distribution of couplings $J_{ij}$, neither the solution of
the matching problem nor the mapping back to spin configurations needs to be unique
\cite{landry:02}: if some edges in the matching problem have the same weight, there
could be different matchings of minimum weight. On the other hand, also the notion of
minimum-weight paths on the original square lattice is not necessarily unique, and
more than one path between two frustrated plaquettes could be of minimal weight.
Finally, each configuration of energy strings corresponds to two different spin
states, related by spin inversion. This generally leads to a large ground-state
degeneracy for discrete and rational distributions $P(\{J_{ij}\})$, but a unique
ground state, e.g., for the Gaussian case \cite{amoruso:03a}. For the complete graph
${\cal G}$, the time complexity of Edmonds' implementation would be O($|{\cal
  F}|^4$), corresponding to O($L^8$) for a $L\times L$ lattice. Although this is
polynomial, further improvements are highly desirable to reduce the rather large
exponent and enable treatment of reasonably sized problem instances. This can be
achieved by a thinning of the complete graph ${\cal G}$: a matching of frustrated
plaquettes in the ground state becomes more and more unlikely with increasing weight
of the path connecting them, and such large-weight edges can consequently be
disregarded. Suitable cutoff parameters depend on the distribution of $J_{ij}$, and
have to be tested thoroughly. For the implementation used here, cutoffs at fixed
maximum path weight and conditions on the minimum vertex degree in the matching graph
have been employed with comparable success.

\subsection{Embedded matching for continuous spins\label{sec:embedded_matching}}

\begin{figure}[tb]
  \centering
  \includegraphics[clip=true,keepaspectratio=true,width=5cm]{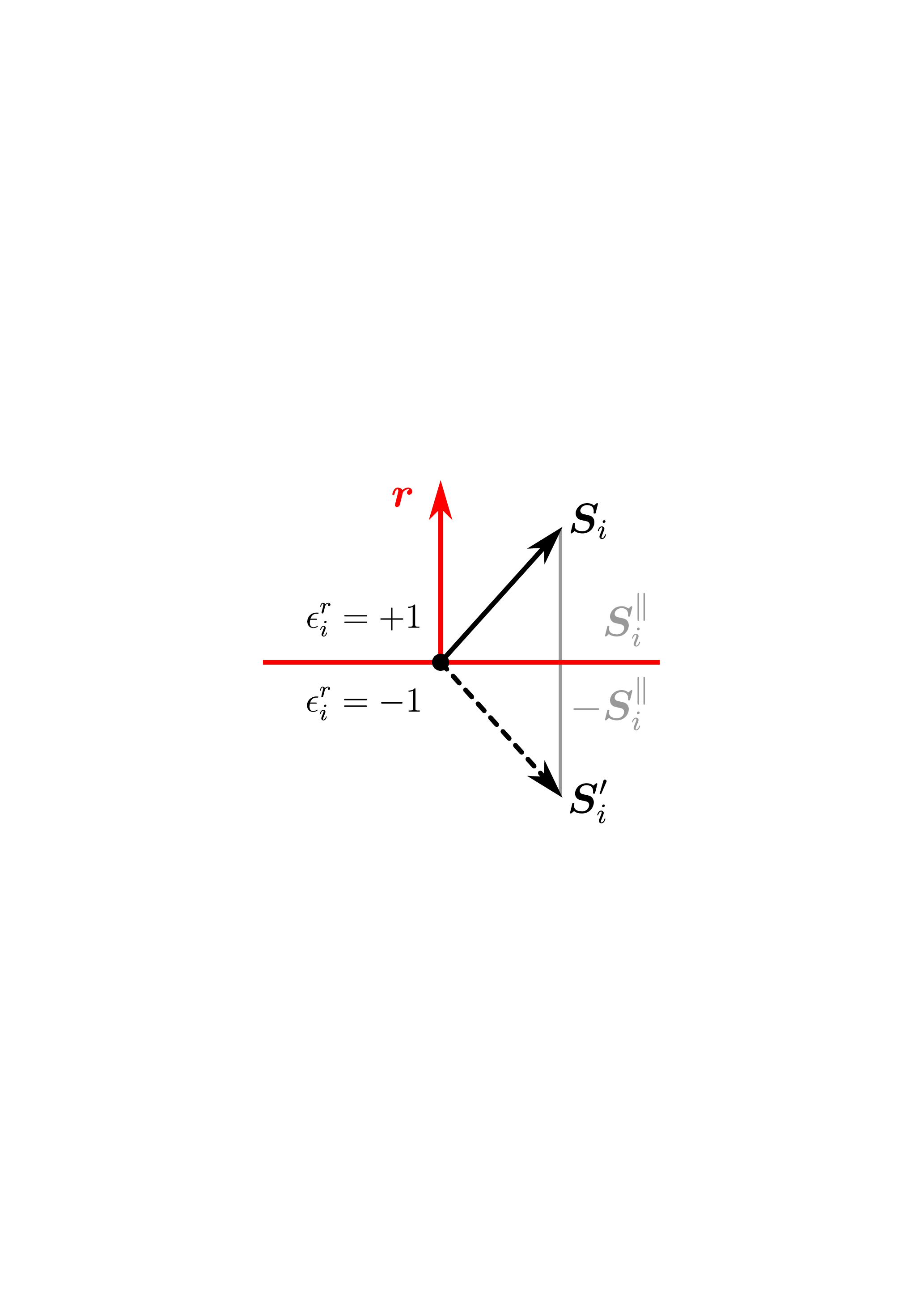}
  \caption
  {(Color online) Embedding of Ising spins into the continuous rotators $\bm{S}_i$
    via decomposition with respect to a direction $\bm{r}$ in spin space.}
  \label{fig:embedding}
\end{figure}

For continuous spins, the notion of plaquette frustration stays meaningful, since it
is a property of the bond configuration only. The transformation to a matching
problem, however, is restricted to discrete Ising spins. To leverage the tractability
of the Ising case for the treatment of continuous-spin systems, an {\em embedding\/}
of Ising spins into the continuous rotators is employed. To this end, consider an
arbitrary direction $\bm{r}$, $|\bm{r}| = 1$, in spin space common to all lattice
sites, and decompose the O($n$) spins $\bm{S}_i$ of Eq.~(\ref{eq:EA_model}) as
$\bm{S}_i = \bm{S}_i^\parallel + \bm{S}_i^\perp = (\bm{S}_i\cdot\bm{r})\bm{r} +
\bm{S}_i^\perp$, cf.\ the illustration in Fig.~\ref{fig:embedding}. This induces a
decomposition ${\cal H} = {\cal H}^{r,\parallel} + {\cal H}^{r,\perp}$ with
\begin{equation}
  \label{eq:embedded_hamiltonian}
  {\cal H}^{r,\parallel} = -\sum_{\langle i,j\rangle}\tilde{J}^r_{ij}\,\epsilon_i^r
  \epsilon_j^r,
\end{equation}
where $\epsilon_i^r = \mathrm{sign}(\bm{S}_i\cdot\bm{r})$, and where the effective
couplings $\tilde{J}^r_{ij}$ are given by
\begin{equation}
  \label{eq:embedded_couplings}
  \tilde{J}_{ij}^r = J_{ij}|\bm{S}_i\cdot\bm{r}||\bm{S}_j\cdot\bm{r}|.
\end{equation}
Hence, with respect to reflections of the spins $\bm{S}_i$ along the plane defined by
$\bm{r}$, the signs $\epsilon_i^r$ are Ising variables (cf.\
Fig.~\ref{fig:embedding}), and the embedded Hamiltonian
(\ref{eq:embedded_hamiltonian}) is that of an Ising model. Note that with respect to
these reflections $\bm{S}_i \mapsto \bm{S}_i-2(\bm{S}_i\cdot\bm{r})\bm{r}$, the
perpendicular part ${\cal H}^{r,\perp}$ is invariant and thus does not contribute to
the embedded dynamics. A similar embedding of Ising variables has been used to
formulate a cluster-update Monte Carlo algorithm for continuous spins
\cite{wolff:89a}.

Updating the effective Ising variables $\epsilon_i^r \mapsto -\epsilon_i^r$ via plane
reflections $\bm{S}_i \mapsto \bm{S}_i-2(\bm{S}_i\cdot\bm{r})\bm{r}$, a ground state
of the Hamiltonian ${\cal H}^{r,\parallel}$ of (\ref{eq:embedded_hamiltonian}) can be
found, for instance, using the transformation to a matching problem outlined above.
Note that since $\sign \tilde{J}_{ij}^r = \sign J_{ij}$ according to
Eq.~(\ref{eq:embedded_couplings}), the frustration function $\Phi_{\square_n}$ does
not depend on the embedding direction $\bm{r}$, and only the weights of the matching
graph ${\cal G}$ must be updated for each embedded matching computation. The
rotational symmetry of the O($n$) Hamiltonian (\ref{eq:EA_model}) can then be
recovered by a random sampling over different embedding directions $\bm{r}$. This
leads to the following algorithm:

\vspace*{2ex}
\begin{algorithmic}[1]
\Procedure{EmbeddedMatching}{$\{\bm{S}_i\}, {\cal I}$}
\For{$j \gets 1, {\cal I}$}
  \State choose random direction $\bm{r}_j$
  \State determine ground state $\hat{\epsilon}_i^{r_j}$
         of ${\cal H}^{r_j,\parallel}(\epsilon_i^{r_j})$
  \For{$i \gets 1, L^2$}
     \If{$\epsilon_i^{r_j} \ne \hat{\epsilon}_i^{r_j}$}
       \State $\bm{S}_i \gets \bm{S}_i-2(\bm{S}_i\cdot\bm{r}_j)\bm{r}_j$
     \EndIf
  \EndFor
\EndFor
\EndProcedure
\end{algorithmic}
\vspace*{2ex}

For each direction $\bm{r}$, it holds that ${\cal H} = {\cal
  H}^{r,\parallel}(\{\epsilon_i^r\}) + {\cal H}^{r,\perp} \ge {\cal
  H}^{r,\parallel}(\{\hat{\epsilon}_i^r\}) + {\cal H}^{r,\perp}$, where
$\hat{\epsilon}_i^r$ is the ground state configuration of the embedded Ising model.
Consequently, the embedded matching procedure corresponds to a strictly downhill
minimization approach. If ${\cal H}$ is in a ground state, ${\cal
  H}^{r,\parallel}(\{\epsilon_i^r\}) = {\cal H} -{\cal H}^{r,\perp}$ must be in a
ground state for each $\bm{r}$ as well. Conversely, however, ${\cal
  H}^{r,\parallel}(\{\epsilon_i^r\})$ being in a ground state for each $\bm{r}$ does
not guarantee global minimum energy for the full ${\cal H}$. This is due to the fact
that the embedded couplings $\tilde{J}_{ij}^r$ of (\ref{eq:embedded_couplings})
depend on the spin configuration $\{\bm{S}_i\}$ and hence on the history of previous
embedding matching runs. As a consequence, the dynamics of $\{\bm{S}_i\}$ induced by
the embedded matching procedure has metastable states \footnote{It is easy to see,
  for instance, that a pure Ising ground state $\bm{S}_i = \pm
  (1,0,\ldots)^\mathrm{T}$, $i=1,\ldots,L^2$ is invariant under the embedded matching
  algorithm as well as the local spin quench (\ref{eq:internal_field}).}. The number
of metastable states is far less, however, than for the local spin-quench approach of
Eq.~(\ref{eq:internal_field}), since for each direction $\bm{r}$ a {\em global\/}
minimum is found: while ${\cal H}^{r_i,\parallel}$ converges to a locally spin-flip
stable state (otherwise a direction $\bm{r}$ could be found, for which embedded Ising
minimization would lead to reflections of one or more spins), not every such locally
stable state is metastable with respect to the embedded matching procedure (because
the embedded Ising system for some direction $\bm{r}$ could be metastable instead of
globally optimal).

\begin{figure}[t]
  \centering
  \includegraphics[keepaspectratio=true,scale=0.75,trim=45 48 75 78]{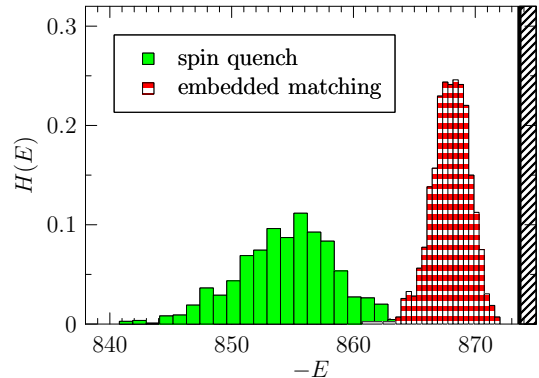}
  \caption
  {(Color online) Histograms of energies of metastable states of a $24\times24$
    sample of the $\pm J$ {\em XY\/} spin glass on the square lattice as found from
    $1\,000$ independent runs of the embedded matching technique compared to results
    of a local spin quench. The onset of the hatched area to the right indicates the
    ground-state energy of this sample. \label{fig:embedded_matching}}
\end{figure}

It is found numerically that the sequence $\{E_i\}$ of energies of the embedded
matching approach always converges. As a consequence of the dependency of
$\tilde{J}_{ij}^r$ of (\ref{eq:embedded_couplings}) on $\{\bm{S}_i\}$, the limit
$E_\infty$ depends on the particular sequence $\{\bm{r}_1,\bm{r}_2,\ldots\}$ of
chosen directions.  Figure~\ref{fig:embedded_matching} shows a histogram of energies
found from the embedded matching approach with a number of different random starting
configurations and different series of embedding directions for a particular
$24\times 24$ sample of the 2D bimodal {\em XY\/} spin glass. For comparison, a
corresponding histogram for the local spin-quench method of
Eq.~(\ref{eq:internal_field}) is also shown. It is apparent that the average energy
of metastable states is lower for the embedded matching technique, and this behavior
is found to survive averaging over the random couplings $\{J_{ij}\}$. Nevertheless,
the probability for converging to a ground state is apparently very small for the
system size considered, cf.\ Fig.~\ref{fig:embedded_matching}. One might speculate
that this shortcoming is connected to the fact that only reflections of spins along a
plane, i.e., improper rotations, are allowed updates in this approach: if the
intermediate, improperly rotated configurations connecting a state to another,
properly rotated state of lower energy have higher energy, they form a barrier which
cannot be overcome by a strictly downhill procedure. The only other transformation
${\cal R}$ besides plane reflections allowing for an Ising type symmetry ${\cal R}^2
= \mathrm{id}$ are point inflections $\bm{S}_i \mapsto -\bm{S}_i$. The embedded
matching technique can be extended to include these transformations. Their inclusion,
however, is not found to remove significantly many barriers, such that this approach
is not further considered here \footnote{Computationally, this enlargement of
  transformations is not very efficient since for inversions $\bm{S}_i \mapsto
  -\bm{S}_i$ the identification of frustrated plaquettes of the embedded Ising model
  depends on the configuration $\{\bm{S}_i\}$ of the O($n$) spins, i.e., one finds
  $\sign \tilde{J}_{ij}^r \neq \sign J_{ij}$ in general.}.

\section{Bond-energy difference crossover and genetic matching\label{sec:genetic}}

Although an important improvement over the local spin quench approach
(\ref{eq:internal_field}), embedded matching alone is not sufficient for reliably
finding ground states. Further advances are possible by an understanding of the
structure of metastable states exploited in a suitably tailored genetic algorithm.

\subsection{Rigidity and domain structure\label{sec:BED}}

To understand the mechanism of metastability in the embedded matching approach and
develop a strategy for overcoming it, one needs to take into account some features of
the low-temperature phase of spin glasses. While there is no long-range order, the
freezing of spin orientations corresponds to some short-range order, expressed in a
non-zero range of correlations \cite{binder:86a}. Consequently, at low temperatures
spins are rather {\em rigidly\/} locked together locally, and their orientation can
only be changed (at low, but generally non-zero energies) by a rigid O($n$) rotation
of a cluster of spins \cite{henley:84a}. Therefore, the manifold of internal states
(i.e., the parameter space of the relevant order parameter) is described by the full
orthogonal group O($n$), in contrast to the case of homogeneous magnets, where the
global magnetization confines the internal states to the quotient space
$\mathrm{SO}(n)/\mathrm{SO}(n-1) \simeq S^n$, i.e., an $n$-dimensional unit sphere
\cite{mermin:79,toulouse:79a}. Such spin clusters hence behave like solid
$n$-dimensional bodies. Note, however, that their O($n$) rotation is not in general a
zero mode, but a low-energy excitation. The existence of such clusters could recently
be explicitly revealed utilizing the genetic embedded matching approach for the
planar spin glass in two dimensions \cite{weigel:05f,weigel:06c}. This symmetry also
determines the topologically stable defects in spin glasses: as in ferromagnets, they
are determined by the homotopy groups of the internal space, here O($n$). For planar
rotators, for instance, in addition to vortices (resp.\ vortex lines) also present in
the homogeneous case, this framework predicts domain walls, which can be directly
observed in form of chiral walls for the (two-dimensional) {\em XY\/} spin glass
\cite{weigel:06,weigel:06c}.  Consequently, some important classes of low-energy
excitations in continuous spin glasses are:
\begin{enumerate}
\item Rigid O($n$) rotations of spin domains.
\item Topological defects: domain walls, vortices etc.
\item Smooth, spin-wave excitations.
\end{enumerate}
In the context of ground-state searches, spin waves can be easily removed by local
relaxation techniques (see the discussion below in Sec~\ref{sec:genetic_matching}).
Some of the topological defects, such as domain walls, can be composed out of a
sequence of domain rotations, such that I concentrate on this first type of
excitation here. Note that the given classification is not meant to be exhaustive,
i.e., it does not express a prejudice as to whether the asymptotic low-energy
excitations in spin glasses are droplets \cite{fisher:86,bray:87a}, mean-field like
extended defects \cite{mezard:84} or ``sponges'' \cite{houdayer:00a}, for instance.
It is, instead, only used as a guideline for the identification of appropriate
metavariables in the formulation of an efficient ground-state search heuristic.

\begin{figure}[t]
  \centering
  \includegraphics[keepaspectratio=true,scale=0.75,trim=75 48 75 78]{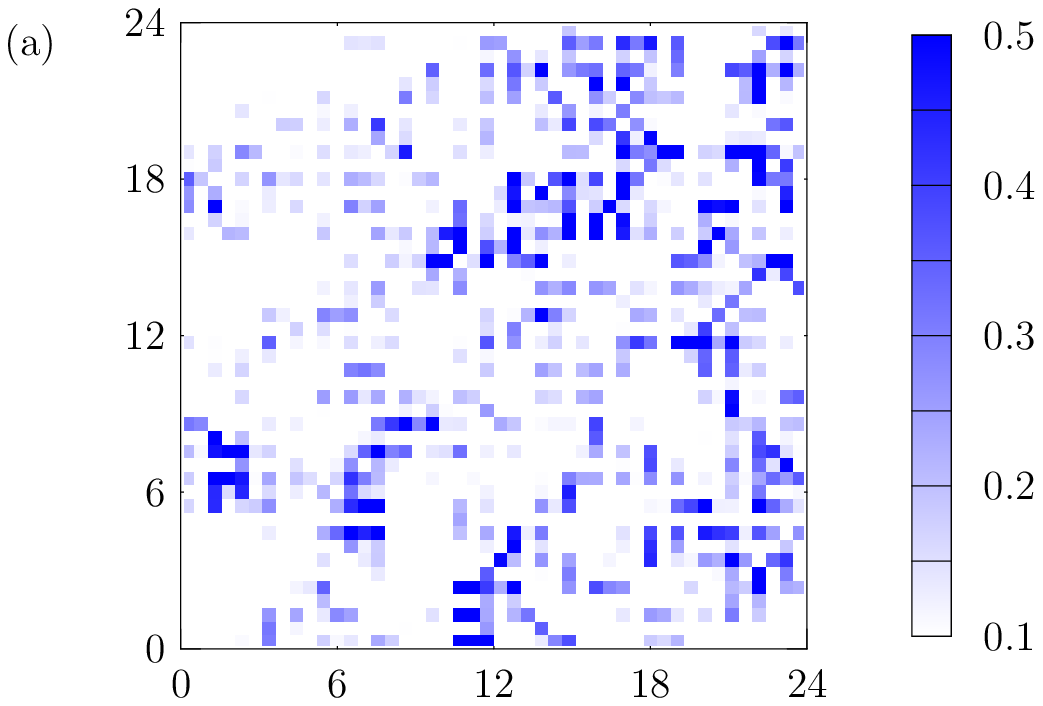}
  \includegraphics[keepaspectratio=true,scale=0.75,trim=75 48 75 78]{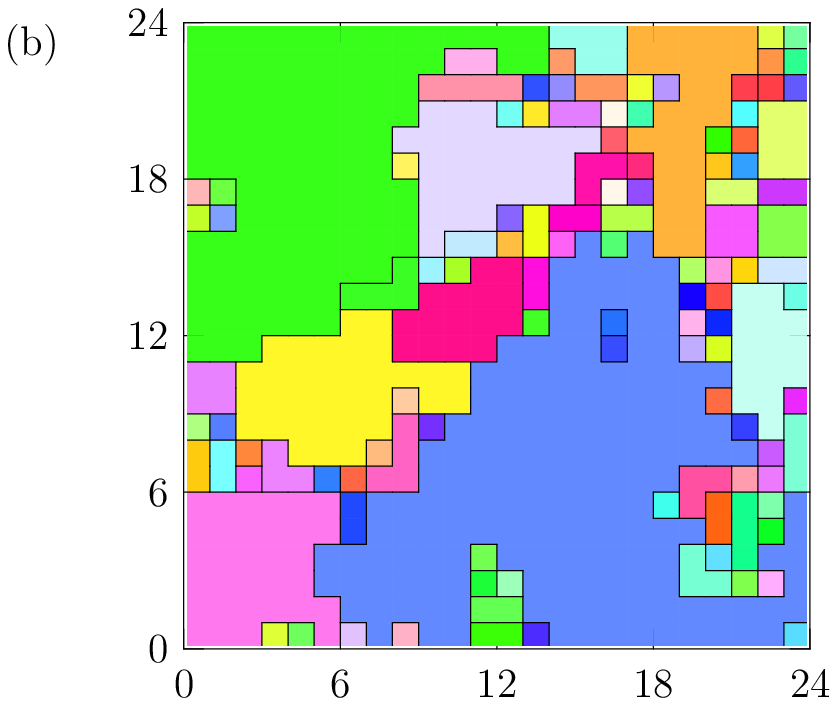}
  \caption
  {(Color online) Relative domain decomposition of metastable states of the
    two-dimensional {\em XY\/} spin glass. (a) Density plot of bond energy
    differences (\ref{eq:bed}) for two metastable states of the embedded matching
    method for a given $24\times 24$ disorder sample. (b) Cluster decomposition
    resulting from the Hoshen-Kopelman algorithm with clustering rule
    (\ref{eq:bed_cluster}) and cutoff parameter $\kappa^{\alpha,\beta} = 0.3
    \sigma_\mathrm{BED}^{\alpha,\beta} \approx 0.05$.\label{fig:BED}}
\end{figure}

Given that typical metastable states differ by rigid O($n$) rotations of domains, an
explicit implementation of such rotations in an optimization heuristic enables it to
perform a search directly on the space of metastable states. Direct inspection of the
transformations connecting metastable states in continuous-spin glasses show that
this domain structure indeed is a valid description, see
Refs.~\cite{henley:84b,weigel:06,weigel:06c}. Note that the concept of such domains
is a relative one, i.e., the domain decomposition of a metastable configuration can
only be determined with respect to other metastable states. In particular, a domain
decomposition may be performed for a {\em pair\/} of configurations, identifying
the set of O($n$) rotations mapping them onto each other. This might be achieved by
determining, by a singular-value decomposition, locally averaged rotation matrices
connecting the configurations \cite{henley:84b,weigel:06c}. Here, a (computationally)
simpler approach is chosen by noting that the O($n$) symmetry of the Hamiltonian
(\ref{eq:EA_model}) ensures that for each such domain all energies $E_{ij} = -
J_{ij}\, \bm{S}_i\cdot\bm{S}_j$ of bonds in the interior are invariant, whereas the
energies carried by bonds crossing the domain boundary change due to local mismatches
of the surface spins of the rotated domain and its environment. Some of these changes
will be of spin-wave type and hence will have been relaxed away once metastability
has been reached again.  Some differences, however, remain, giving a handle on domain
identification. Consequently, for a pair ($\alpha, \beta$) of metastable
configurations one might consider the {\em bond energy differences\/} (BEDs)
\begin{equation}
  \label{eq:bed}
  |\Delta E_{ij}^{\alpha,\beta}| = \left|J_{ij}\left(\bm{S}_i^\alpha\cdot\bm{S}_j^\alpha
      -\bm{S}_i^\beta\cdot\bm{S}_j^\beta\right)\right|.
\end{equation}
The distribution of BEDs is depicted for a pair of metastable states of the 2D {\em
  XY\/} spin glass in Fig.~\ref{fig:BED}(a), showing clear structures of rigid
domains. Defining a domain boundary by a BED exceeding a threshold value, i.e.,
\begin{equation}
  \label{eq:bed_cluster}
  |\Delta E_{ij}^{\alpha,\beta}| > \kappa^{\alpha,\beta},
\end{equation}
a domain decomposition can be performed, for instance, with the Hoshen-Kopelman
algorithm \cite{hoshen:79}. This is illustrated in Fig.~\ref{fig:BED}(b) for the BEDs
of Fig.~\ref{fig:BED}(a). The cutoff parameter $\kappa^{\alpha,\beta}$ is chosen here
of the order of the total variation of BEDs over the whole sample, i.e., proportional
to the standard deviation $\sigma_\mathrm{BED}^{\alpha,\beta}$, to accommodate
differences between disorder realizations as well as metastable states of largely
varying energies. Since these domains are merely utilized for a more efficient
ground-state search, I do not have to bother here with the question of whether there
is a precisely defined characteristic length associated with such domains
\cite{henley:84b}, independent of the size of the system, as having
$\kappa^{\alpha,\beta} \sim \sigma_\mathrm{BED}^{\alpha,\beta}$ detects the length(s)
appropriate to the sample at hand automatically.

\subsection{Genetic matching\label{sec:genetic_matching}}

Embedded matching in combination with domain decomposition of configurations with BED
clustering allows optimization directly on the space of metastable states of the
embedded matching method. This already corresponds to an enormous reduction in the
size of the phase space. The meta-optimization on metastable states is performed here
utilizing a hybrid genetic algorithm, although variants based on other global
optimization strategies such as simulated annealing are conceivable as well. Genetic
algorithms \cite{michalewicz:96} mimic natural evolution by maintaining a population
of candidate solutions, which is evolved in generations by a process involving the
crossover and mutation of solutions followed by a selection of members with higher
fitness, i.e., lower energy for the case of ground-state computations considered
here. In the canonical form of genetic algorithm, solutions are represented by bit
strings in a binary representation and crossover and mutation correspond to the
random exchange of bits between solutions and random bit flips, respectively
\cite{michalewicz:96}. In this form, genetic algorithms have been applied to the
Ising spin glass, but, unless for very small systems, true ground states could not be
found with high reliability \cite{sutton:94,gropengiesser:95}. Only hybrids combining
genetic crossover with some downhill optimization procedure such as local spin flips
or the ``cluster-exact approximation'' \cite{hartmann:96a}, restricting the search
space to metastable states, led to more successful approaches
\cite{pal:96a,hartmann:99a}.

Although widely and successfully employed, there is no theoretically sound framework
for designing efficient genetic algorithms \cite{michalewicz:96}, such that their
construction rests on heuristic strategies and additional insight specific to the
problem at hand. Generally, one strives to achieve a balance between fast convergence
to an optimum answer and the upholding of genetic diversity throughout the
``evolution'' (which, in turn, tends to slow down convergence), in order not to miss
the global optimum. The choice of crossover operation appears to be most important in
this context.  In the present work, instead of randomly exchanging single spins, the
BED cluster decomposition is employed to exchange domains of rigid spins between
solutions.  This has the important advantage of retaining the high degree of
optimization already found from the embedded matching technique {\em inside\/} of the
domains and directly operating on the space of variables relevant for the
construction of metastable states. The domain decomposition can be performed directly
with the ``parent'' configurations to be combined (``diadic'' crossover) or,
alternatively, by using a third, ``mask'' configuration from the population used only
for the domain decomposition. The latter, ``triadic'' crossover is used here, similar
to the technique suggested in Ref.~\cite{pal:95} for Ising spins, since it is found
to perform slightly better for continuous spins.  Genetic diversity is strengthened
by restricting the selection of parents to be combined to neighbors after the
population has been arranged in a linear ring \cite{pal:95}.  This introduces some
degree of geometric ``locality'' of the population and allows good solutions to be
refined independently in different areas of the configuration set.  Previous
approaches \cite{pal:96a,hartmann:99a,palassini:99a} have used a fixed total number
of crossover operations per member of the initial population, followed by a halving
of the population by elimination of the higher-energy solution of each pair of
neighboring configurations, and then a re-iteration of the remaining population.
This reduces the total effort by removing unpromising solutions and bringing distant
parts of the ``ring'' of solutions closer to each other in later stages of the
optimization. For the time being, I will adopt this technique here as well. A more
efficient variant, geared at the detection of hard samples, is presented below in
Sec.~\ref{sec:hardness}. In total, the resulting {\em genetic embedded matching\/}
(GEM) algorithm proceeds as follows:

\vspace*{2ex}
\begin{algorithmic}[1]
\Procedure{GEM}{${\cal S}, {\cal C}, {\cal P}, {\cal I}, {\cal L}$}
\State initialize $\{\bm{S}_i^k\}$, $k=1,\ldots,{\cal S}$ randomly
\State $s \gets {\cal S}$
\While{$s > 4$}
  \For{$c \gets 1, {\cal C}\times s$}
    \State randomly select pair $(\alpha,\beta = \alpha+1)$ of confs.
    \State randomly select mask conf.\ $\gamma \ne \alpha, \beta$
    \State $\{\hat{\bm{S}}_i^\alpha, \hat{\bm{S}}_i^\beta\} = $ 
           \Call{BEDCrossover}{$\alpha$, $\beta$, $\gamma$}
    \State \Call{Mutate}{$\{\hat{\bm{S}}_i^\alpha\}, {\cal P}$},
           \Call{Mutate}{$\{\hat{\bm{S}}_i^\beta\}, {\cal P}$}
    \State \Call{EmbeddedMatching}{$\{\hat{\bm{S}}_i^\alpha\}, {\cal I}$}
    \State \Call{EmbeddedMatching}{$\{\hat{\bm{S}}_i^\beta\}, {\cal I}$}
    \For{$j \gets 1, {\cal L}$}
      \For{$i \gets 1, L^2$}
        \State $\hat{\bm{S}}_i^\alpha \gets \hat{\bm{h}}_i^\alpha/|\hat{\bm{h}}_i^\alpha|$  
        \State $\hat{\bm{S}}_i^\beta \gets \hat{\bm{h}}_i^\beta/|\hat{\bm{h}}_i^\beta|$  
      \EndFor
    \EndFor
    \If{${\cal H}(\{\hat{\bm{S}}_i^\alpha\}) < {\cal H}(\{\bm{S}_i^\alpha\})$}
      \State $\bm{S}_i^\alpha \gets \hat{\bm{S}}_i^\alpha$, $i=1,\ldots,L^2$ 
    \EndIf
    \If{${\cal H}(\{\hat{\bm{S}}_i^\beta\}) < {\cal H}(\{\bm{S}_i^\beta\})$}
      \State $\bm{S}_i^\beta \gets \hat{\bm{S}}_i^\beta$, $i=1,\ldots,L^2$ 
    \EndIf
  \EndFor
  \ForAll{distinct pairs ($\alpha, \alpha+1)$}
    \If{${\cal H}(\{\bm{S}_i^\alpha\}) < {\cal
        H}(\{\bm{S}_i^{\alpha+1}\})$}
      \State remove configuration $\alpha+1$
    \Else
      \State remove configuration $\alpha$
    \EndIf
    \State $s \gets s-1$
  \EndFor
\EndWhile
\State output (best of) remaining configurations
\EndProcedure
\end{algorithmic}
\vspace*{2ex}

The {\sc BEDCrossover} operation performs the BED domain decomposition of
Sec.~\ref{sec:BED} with respect to the mask and, for each domain, either swaps all
spins between the parents $\alpha$ and $\beta$, copies domains $\alpha \mapsto \beta$
or $\beta \mapsto \alpha$, or leaves the domain invariant, with all possibilities
occurring randomly at equal proportions. Mutations are performed by randomly choosing
new spin orientations with a probability ${\cal P}$. The resulting offspring are
optimized using ${\cal I}$ iterations of {\sc EmbeddedMatching} from
Sec.~\ref{sec:embedded_matching}, followed by ${\cal L}$ iterations of a local spin
quench. The latter is useful since, close to a minimum where only spin-wave
excitations are left, both approaches converge to the same state, but the spin quench
is much faster computationally. Lower-energy offspring then replace their parents.
In the implementation used, each offspring is only compared to the morphologically
closer parent, i.e., the one with a larger optimized scalar overlap
$\hat{q}^{\alpha\beta} = \max_{R\in\mathrm{O}(n)}\sum_{\sigma,\tau}
q_{\sigma\tau}^{\alpha\beta} R_{\sigma\tau}^{\alpha\beta}$, where
$R_{\sigma\tau}^{\alpha\beta}$ denotes the corresponding global rotation matrix, and
\begin{equation}
  \label{eq:overlap_matrix}
  q_{\sigma\tau}^{\alpha\beta} = \frac{1}{L^2} \sum_i S_{i\sigma}^{\alpha} S_{i\tau}^{\beta},
\end{equation}
is the matrix of overlaps. This maximization is performed by a singular value
decomposition to diagonalize $q_{\sigma\tau}^{\alpha\beta}$, in which case
$\hat{q}^{\alpha\beta}$ is just the trace of the resulting diagonal
$q_{\sigma\tau}^{\alpha\beta}$ \cite{henley:84b,weigel:06c}. This form of replacement
restriction helps to maximize genetic diversity \cite{pal:96a}. After $s{\cal C}$
crossovers, the higher energy instance of each adjacent pair of configurations is
discarded, thus halving the population. The complete process is repeated until at
most four configurations are left, which form the result of a run.

As will be shown in the next Section, this combination of techniques in the genetic
embedded matching method allows for the determination of (numerically) exact
ground-states of reasonably large continuous-spin systems in 2D with high
reliability.

\section{Performance\label{sec:performance}}

Using probabilistic methods for ground-state searches, special care is needed to
ensure that true ground states are found. Since for ${\cal NP}$ hard optimization
problems the decision variant is ${\cal NP}$ complete, there is no way of definitely
distinguishing a metastable from a ground state short of an exact solution of the
instance.  A general probabilistic approach of ``quality assurance'' for the GEM
method is outlined and applied to the 2D {\em XY\/} spin glass in
Sec.~\ref{sec:probabilistics}. In some dynamical approaches, such as local spin-flip
Monte Carlo simulations, the specific hardness of a sample shows up in the behavior
of autocorrelation times, to which a simulation run can in principle react
dynamically by increasing the simulation time accordingly. Below in
Sec.~\ref{sec:hardness}, it is discussed whether a similar heuristic for detecting
hard samples can be applied in the GEM approach.

\subsection{Performance and comparison to simulated annealing}

\begin{figure}[t]
  \centering
  \includegraphics[keepaspectratio=true,scale=0.75,trim=45 48 75 78]{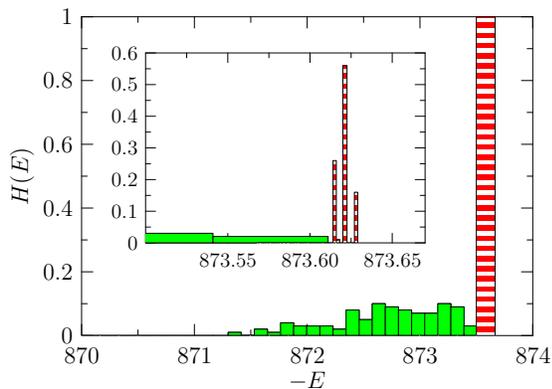}
  \caption
  {(Color online) Histograms of energies found for a $24\times 24$ sample of the
    bimodal {\em XY\/} spin glass from $100$ runs of the genetic embedded matching
    approach with a population of size ${\cal S} = 64$ (hatched bars, average runtime
    $700$ s on a Pentium IV 2.8 GHz) as compared to simulated annealing runs with a
    total of about $2\times 10^7$ Monte Carlo sweeps (solid bars, average runtime
    $4\,000$ s). The disorder realization is identical to the one considered in
    Fig.~\ref{fig:embedded_matching}.\label{fig:GEM_anneal_histo}}
\end{figure}

Local spin quenches according to Eq.~(\ref{eq:internal_field}) yield states in a
broad range of energies, cf.\ Fig.~\ref{fig:embedded_matching}. For ascribing the
ability to find ground states to a stochastic method one would, instead, require that
states of exactly the same energy (up to machine precision) are found in a sizable
fraction $p_s$ of attempts (with $p_s = 95\%$, for instance) {\em and\/} that no
states of lower energy can be found with runs of largely increased effort or
utilizing other optimization techniques. As is evident from
Fig.~\ref{fig:embedded_matching}, this also cannot be said of the embedded matching
approach alone. Figure~\ref{fig:GEM_anneal_histo} shows the minimum energies found in
repeated runs of the GEM technique for the bimodal {\em XY\/} spin glass in 2D with a
randomly picked disorder realization of size $24\times 24$ (which is identical to the
realization used in Fig.~\ref{fig:embedded_matching}) and a starting population size
${\cal S} = 64$. For comparison, this Figure also shows the histogram of repeated
runs of an extensive simulated annealing \cite{kirkpatrick:83,kirkpatrick:84}
computation with an exponential temperature protocol (a linear protocol yields very
similar results) and a total number of about $2\times10^7$ lattice sweeps of single
spin flips per run, leading to an about sixfold runtime as compared to the GEM
computations. As is seen, the GEM runs result in clearly lower energies than the
simulated annealing.  Additionally, the latter still show a sizable spread of the
energies found, whereas the GEM technique appears to yield states of the same energy.
Only on going to much higher energy resolution, the GEM results are resolved into a
small number of distinct peaks cumulated from runs yielding identical energy up to
(or close to) machine precision, cf.\ the inset of Fig.~\ref{fig:GEM_anneal_histo}. A
fourfold increase of the starting population to ${\cal S} = 256$ leads to a
convergence of all 100 runs to the lowest-energy peak to the right in the inset of
Fig.~\ref{fig:GEM_anneal_histo}. No further increases of the population size up to
${\cal S} = 1024$ lead to lower energies such that, with high confidence, this peak
corresponds to the true ground-state energy of the system.  Consequently, it can be
said that runs with ${\cal S} = 64$ have a probability of about $p_s = 16\%$ of
leading to a ground-state. Methods for guaranteeing high reliability of finding
ground states over the distribution of disorder realizations are discussed below in
Secs.~\ref{sec:probabilistics} and \ref{sec:hardness}.

Although, for the given disorder realization, the GEM technique appears able to find
ground states and clearly outperforms the simulated annealing approach, variations in
the ``hardness'' of different replica in the random couplings are known to be large
(see, e.g., Refs.~\cite{berg:98a,alder:04}), and the corresponding variation in the
efficiency of the methods should be taken into account. Since the commonly considered
distributions $P(J_{ij})$ contain the ferromagnetic lattice with $J_{ij} = J_0 > 0$,
which is trivially handled by either optimization method, the behavior of interest
can only be either that for the worst case, which is, however, difficult to assess
for the spin-glass model considered, or rather the {\em average\/} performance for
the disorder distribution at hand. As a first step in this analysis, I considered the
convergence of the average minimum energy observed with the computational effort
invested. For simulated annealing with Metropolis acceptance rule, it is known that
with logarithmically slow cooling, ground states will be found in finite (but, of
course, very large) time \cite{geman:84}. Since this cooling schedule is impractical,
however, exponential or power-law cooling curves are used instead
\cite{kirkpatrick:83,kirkpatrick:84}. The possibility of different acceptance rules
complicates things further, and it is naturally impossible to benchmark against all
these variants. I restrict myself here to the probably most commonly used exponential
protocol. The asymptotic form of energy convergence in simulated annealing of
spin-glass systems has been the topic of some debate in the past
\cite{grest:86,huse:86}. Numerically, a power-law convergence,
\begin{equation}
  \label{eq:anneal_decay}
  \langle E(T)\rangle_J \sim E_\infty + A_E T^{-\zeta},
\end{equation}
\begin{figure}[t]
  \centering
  \includegraphics[keepaspectratio=true,scale=0.75,trim=45 48 75 78]{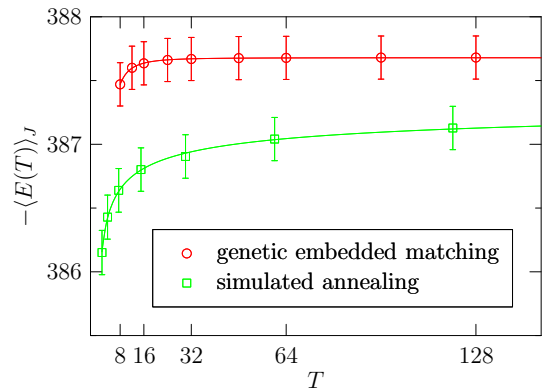}
  \caption
  {(Color online) Average minimum energy $\langle E(T)\rangle_J$ for $1\,000$ samples
    of size $16\times 16$ of the bimodal 2D {\em XY\/} spin glass found from GEM and
    simulated annealing runs with a total runtime $T$ (in re-scaled units).
    \label{fig:anneal_fits}}
\end{figure}
for large cooling times $T$ (i.e., the total number of Monte Carlo sweeps) was found
for the 2D Ising spin glass, while, on the contrary, logarithmic convergence was
observed for the 3D variant \cite{grest:86}. On the other hand, on the basis of
modeling (Ising) spin glasses as sets of weakly interacting two-level systems, it was
conjectured that the logarithmic form would be universal \cite{huse:86}. Here,
$\langle\cdot\rangle_J$ denotes the average over disorder.
Figure~\ref{fig:anneal_fits} shows the average minimum energy found from simulated
annealing of $1\,000$ samples of size $16\times 16$ of the 2D {\em XY\/} spin glass
for annealing times between $T=50\,000$ and $T=3.2\times 10^6$ sweeps, compared to
the energies found from GEM runs with population sizes ${\cal S} \propto T$ between
$8$ and $128$ configurations. The abscissa for the simulated annealing data has been
linearly rescaled to result in equal runtimes for both approaches on a Pentium IV 2.8
GHz (both algorithms scale linearly in $T$). The data from simulated annealing can be
fitted with reasonable quality to the power-law form (\ref{eq:anneal_decay}),
yielding a decay exponent $\zeta = 0.33(36)$, whereas a logarithmic form does not
adequately describe the data.  This is comparable to the value $\zeta = 0.25$ found
for the 2D Ising spin glass in Ref.~\cite{grest:86}. Note that the extrapolated
asymptotic ground-state energy $E_\infty = -387.45(76)$ is compatible statistically
with the value found from GEM runs already for the smallest population size ${\cal S}
= 8$ considered. In fact, the GEM data are constant within statistical errors for
${\cal S} \ge 32$. The variation of energies found from the GEM technique can also be
described by (\ref{eq:anneal_decay}), resulting in $\zeta = 2.3(48)$ (the large
statistical error results from the only minute variation of $\langle E\rangle$
observed).

The GEM algorithm as presented in Sec.~\ref{sec:genetic_matching} involves a number
of parameters which need to be tuned to achieve these good results. Performance
appears to be rather weakly dependent on the mutation rate, and best results are
found for a rate of about ${\cal P} = 2.5\%$. Much more frequent mutation destroys
the relatively good optimization achieved at intermediate stages and decrease the
overall performance. Since the offspring configurations produced by the BED crossover
are still optimal inside of domains, relatively small numbers of embedded matching
and local relaxation steps are found to be sufficient, ${\cal I} = 15$ and ${\cal L}
= 100$ was usually chosen here (cf.\ the pseudocode of the algorithm in
Sec.~\ref{sec:genetic_matching}). The number ${\cal C}$ of crossovers per replica
determines how well the available ``genetic pool'' of original configurations is
explored. Beyond a certain number of crossovers, the population becomes uniform and
further increases do not improve the probability of finding ground states. For
accessible system sizes ${\cal C} = 8$--$16$ is a good choice. The main tuning
parameter of the approach is the initial population size ${\cal S}$, which is changed
to accommodate the variable hardness of different system sizes, models and disorder
realizations. It is the only parameter whose increase ultimately guarantees ground
states to be found. Note that the total number of crossovers is $2{\cal C}({\cal
  S}-4)$ (assuming ${\cal S} = 2^n$) and hence linear in ${\cal S}$. For a given
single realization, computation of true ground states can be guaranteed with high
confidence by tackling the same disorder configuration with largely increased
computational effort (in particular, say, a fourfold increase in population size
${\cal S}$), until no further change in energy is observed. For the random
distribution of configurations to be investigated, however, a more automatic (and
less computationally expensive) approach is required.

\subsection{Probabilistics of successes\label{sec:probabilistics}}

For stochastic optimization methods, arrival at true ground states cannot be
guaranteed.  For given input data in form of the disorder realization and a choice
for the tunable parameters, a ground state is found with the success probability
$p_s(\{J_\mathrm{ij}\}; T)$, where $T$ denotes the relevant parameters. As was
discussed above, by far the most influential parameter for the GEM approach is the
initial population size ${\cal S}$, such that I here restrict considerations to $T =
\{{\cal S}\}$. Full information about the distribution of $p_s$ induced by
$P(\{J_{ij}\})$ and the dependency on the algorithm's parameters would correspond to
a complete understanding of the performance characteristics or generalized
computational complexity \cite{mertens:02}. This computation, however, is impractical
due to the high-dimensional nature of this parameter space: using, e.g., $100$ runs
to estimate $p_s$ for a given set of parameters for $1\,000$ disorder realizations
and $100$ combinations of parameters ${\cal S}, {\cal I},\ldots$ would require $10^7$
ground-state computations for a single system size! From $p_s(\{J_\mathrm{ij}\}; T)$
one could deduce the perhaps most interesting distribution
$T_\mathrm{min}(\{J_\mathrm{ij}\}; p_s)$ of efforts required for a constant success
probability $p_s$.

\begin{figure}[t]
  \centering
  \includegraphics[keepaspectratio=true,scale=0.75,trim=45 48 75 78]{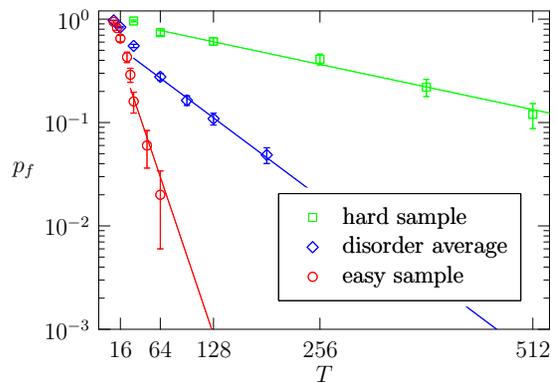}
  \caption
  {(Color online) Estimated failure probability $p_f = 1-p_s$ for the GEM technique
    applied to {\em XY\/} spin-glass samples of size $16\times 16$ as a function of
    the computational effort $T={\cal S}$. The lines show fits of the form
    (\ref{eq:failure_law}) (easy and hard sample) resp.\
    (\ref{eq:average_failure_law}) (disorder average) to the data.
    \label{fig:success_T}}
\end{figure}

Figure~\ref{fig:success_T} shows the failure probabilities $p_f = 1-p_s$ for an
``easy'' and a ``hard'' sample of the 2D bimodal {\em XY\/} spin glass as a function
of the population size ${\cal S} = T$, compared to the average failure rate $p_f$
over $100$ disorder replica. The huge spread in hardness is apparent: while, for
instance, only in 2 out of 100 cases do runs with ${\cal S} = 64$ fail to find a
ground state for the easy sample, $74\%$ of attempts for the hard sample end in a
metastable state. It is therefore not enough to fix the run parameters by considering
one or two randomly chosen configurations. For a description of the functional form
of $p_f(T)$ in Fig.~\ref{fig:success_T}, consider performing $n$ statistically
independent runs of length $T_0$ with failure probability $p_{f,0}$ and picking the
solution of lowest energy as final answer. With this prescription, a ground state is
not being found only if {\em all\/} of the runs fail, and the combined failure
probability is thus
\begin{equation}
  \label{eq:failure_law}
  p_f(T = nT_0) = p_{f,0}^{T/T_0}.
\end{equation}
Due to the locality constraint in choosing parent configurations for crossover,
increasing the initial population size ${\cal S} = T$ has essentially the same effect
as performing independent runs. For sufficiently large $T$,
Eq.~(\ref{eq:failure_law}) is hence an excellent description of $p_f(T)$ for a single
sample. Figure \ref{fig:success_T} shows fits of the form (\ref{eq:failure_law}) to
the data for the ``easy'' and ``hard'' samples. There is only a single fit parameter,
$p_{f,0}^{1/T_0}$, corresponding to a measure of hardness of the sample (with respect
to the GEM technique). The ensemble average $\langle p_f(\{J_{ij}\}; T)\rangle_J$
cannot be expected to follow the same exponential form (\ref{eq:failure_law}) since,
in general, $\langle \exp [\alpha T]\rangle \ne \exp [\langle\alpha\rangle T]$. It is
found, however, that it is well described by the slight generalization
\begin{equation}
  \label{eq:average_failure_law}
  \langle p_f(\{J_{ij}\}; T)\rangle_J = A_p\,p_{f,0}^{T/T_0},
\end{equation}
with an additional amplitude $A_p < 1$, as is apparent from the corresponding fit
also shown in Fig.~\ref{fig:success_T}. Consequently, the average failure probability
decreases more slowly with $T$ than would be expected from the behavior on single
samples. Note that due to the form (\ref{eq:failure_law}) it is not appropriate to
consider the combination $T/p_s$ as the ``computational effort'' of a sample
\cite{alder:04}, since this assumes a linear relation between $p_s$ and $T$.

\begin{figure}[t]
  \centering
  \includegraphics[keepaspectratio=true,scale=0.75,trim=45 48 75 78]{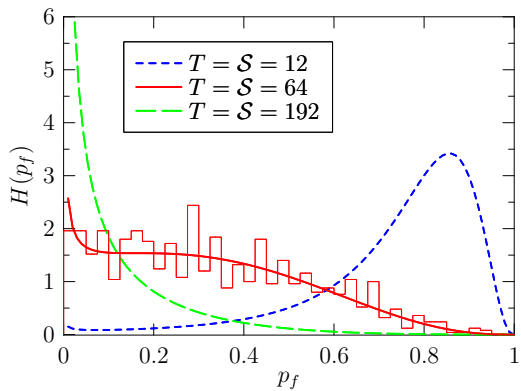}
  \caption
  {(Color online) Histogram $H(p_f)$ of failure probabilities for $1\,000$ disorder
    samples for the 2D {\em XY\/} spin glass as a function of initial population size
    ${\cal S}$. The lines show analytical approximations discussed below in the main
    text.
    \label{fig:success_J}}
\end{figure}

\begin{figure}[t]
  \centering
  \includegraphics[keepaspectratio=true,scale=0.75,trim=45 48 75 78]{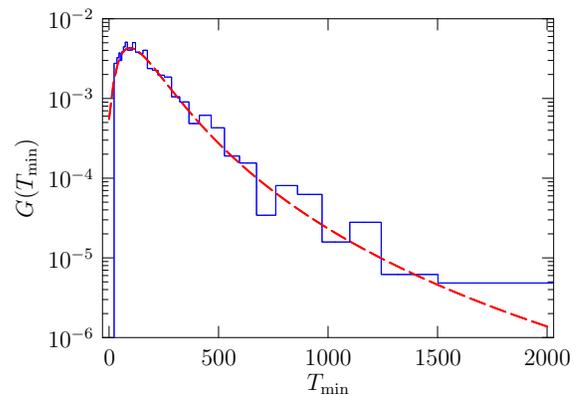}
  \caption
  {(Color online) Estimated probability-density function $G(T_\mathrm{min})$ of
    minimum required runtimes $T_\mathrm{min}$ for $16\times 16$ systems of the 2D
    {\em XY\/} spin glass to achieve success probability $p_s = 95\%$. The dashed
    line shows a fit of the generalized extreme-value distribution (\ref{eq:GEV}) to
    the data.
    \label{fig:GEV}}
\end{figure}

In view of the results of Fig.~\ref{fig:success_T}, it is of interest to investigate
the distribution $H(p_f)$ of failure probabilities over disorder samples. To this
end, the failure probability was sampled by performing $100$ independent runs for
each of $1\,000$ disorder samples with run length $T={\cal S}=64$. The histogram
estimating the probability-density function $H(p_f$) is shown in
Fig.~\ref{fig:success_J}, revealing clearly the breadth of this distribution,
reflecting the large spread in hardness already suggested by the data of
Fig.~\ref{fig:success_T}. From the histogram $H(p_f)$, it is possible by means of
Eq.~(\ref{eq:failure_law}) to recover the distribution of the minimum required
runtimes $T_\mathrm{min}(\{J_\mathrm{ij}\}; p_f)$, corresponding to the distribution
of hardness of samples under the GEM technique: from the estimate of
$p_f(\{J_{ij}\})$ for a disorder sample $\{J_{ij}\}$ at fixed runtime $T$,
Eq.~(\ref{eq:failure_law}) implies
\begin{equation}
  \label{eq:min_runtime}
  T_\mathrm{min}(\{J_\mathrm{ij}\}) = T\,\frac{\ln p_{f,0}}{\ln p_f(\{J_{ij}\})}.
\end{equation}
Figure~\ref{fig:GEV} shows the distribution $G(T_\mathrm{min})$ of minimum runtimes
at failure probability $p_{f,0} = 0.05$ thus resulting from the runs at fixed runtime
$T=64$ presented in Fig.~\ref{fig:success_J}. The breadth of the distribution is
apparent: while the average is around $\langle T_\mathrm{min}\rangle_J \approx 225$,
there is a fat tail with some of the $1\,000$ disorder configurations featuring a
$T_\mathrm{min}$ as large as $2\,000$. Such density functions typically occur when
considering the extrema of large samples drawn from underlying probability
distributions. Asymptotically, the distribution of extremes is known to be universal,
following the form \cite{castillo:88}
\begin{eqnarray}
  \label{eq:GEV}
  G_{\xi,\mu,\sigma}(x) & = & \displaystyle
  \frac{1}{\sigma}\left(1+\xi\frac{x-\mu}{\sigma}\right)^{-1-1/\xi} \nonumber\\
  & & \times \exp\left[-\left(1+\xi\frac{x-\mu}{\sigma}\right)^{-1/\xi}\right],
\end{eqnarray}
where the parameter $\xi$ depends on the tail behavior of the underlying, primary
distribution for large arguments $x$. Depending on $\xi$, this form is known as
Weibull ($\xi < 0$), Gumbel ($\xi \rightarrow 0$) or Fr\'echet ($\xi > 0$)
distribution, respectively. As is seen in Fig.~\ref{fig:GEV}, it fits the data for
$T_\mathrm{min}$ extremely well, resulting in $\xi = 0.270(44)$, $\mu = 115.6(45)$,
and $\sigma = 87.7(39)$ with an excellent quality-of-fit $Q = 0.23$. Similar
distributions of the Fr\'echet type have been found for the tunneling times in
Wang-Landau flat-histogram simulations of the Ising spin glass
\cite{dayal:04,alder:04}. One might speculate on the origin of the occurrence of
extreme-value statistics in hardness measures of spin-glass samples: if, as has been
suggested \cite{huse:86}, a spin-glass sample can be described as a set of $n = n(L)$
weakly interacting two-level systems, it appears plausible that the largest barrier
or the slowest relaxation time determines the hardness of the configuration.  Then,
the hardness would be the maximum or minimum of a sample of size $n$ from the
underlying distribution of two-level systems, asymptotically distributed according to
the extreme-value distribution (\ref{eq:GEV}). In line with this argument, it was
recently observed \cite{bittner:06} that the distribution of relevant barriers in the
Sherrington-Kirkpatrick model follows a Fr\'echet distribution with a value of $\xi
\approx 0.33$, rather similar to the form found here.

Given that $G_{\xi,\mu,\sigma}(T_\mathrm{min})$ describes the GEM data so well, it is
worthwhile to use Eq.~(\ref{eq:failure_law}) to reveal the resulting analytical form
of the distribution of failure probabilities, $H(p_f)$. Following standard
probability theory \cite{feller:68}, density functions transform as
\begin{equation}
  \label{eq:distribution_trafo}
  H(p_f)\,\d p_f = G_{\xi,\mu,\sigma}(T_\mathrm{min}) \frac{\d T_\mathrm{min}}{\d p_f}\,\d p_f,
\end{equation}
which, using Eq.~(\ref{eq:failure_law}), leads to
\begin{equation}
  \label{eq:min_runtime_distrib}
  H(p_f) = -\frac{T\ln p_{f,0}}{p_f(\ln p_f)^2}\,G_{\xi,\mu,\sigma}(T\,\frac{\ln
    p_{f,0}}{\ln p_f}).
\end{equation}
As is seen from the curves in Fig.~\ref{fig:success_J}, this form with the parameter
values $\xi$, $\mu$ and $\sigma$ given above fits the numerical distribution $H(p_f)$
for the same $T = 64$ data perfectly well (i.e., the approach is self-consistent).
Additionally, however, it describes independent runs of different lengths to high
precision and hence the form (\ref{eq:min_runtime_distrib}) is an excellent
description for general runtimes $T$, as indicated by the additional curves in
Fig.~\ref{fig:success_J}. Consequently, the three-parameter family of distributions
(\ref{eq:GEV}) and the limiting distributions derived via Eq.~(\ref{eq:failure_law})
form a complete description of the full probability density $p_f(\{J_\mathrm{ij}\};
T)$.

\begin{figure}[t]
  \centering
  \includegraphics[keepaspectratio=true,scale=0.75,trim=45 48 75 78]{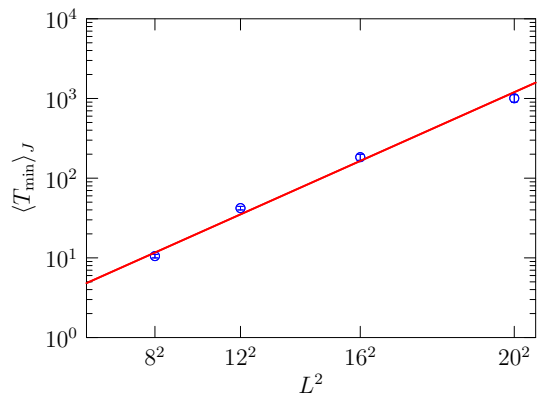}
  \caption
  {(Color online) Average effort $\langle T_\mathrm{min}\rangle_J$ for finding ground
    states with constant success probability $p_s = 95\%$ as a function of lattice
    size $L^2$, estimated from $100$ disorder configurations per size. The line shows
    a fit of the form (\ref{eq:exponential_effort}) to the data.
    \label{fig:exponential}}
\end{figure}

While it certainly would be instructive to extend the analysis of $T_\mathrm{min}$
via the distributions (\ref{eq:GEV}) to a scaling analysis of the fit parameters
$\xi$, $\mu$ and $\sigma$ with lattice size $L$, the huge computational effort would
be disproportional. Instead, I concentrate on the mean required effort $\langle
T_\mathrm{min}\rangle_J$ as a function of system size, averaged over a smaller
disorder sample of only $100$ configurations. These data for failure probability $p_f
= 5\%$ are shown in Fig.~\ref{fig:exponential} together with a fit to the expected
exponential form
\begin{equation}
  \label{eq:exponential_effort}
  \langle T_\mathrm{min}\rangle_J = A_T e^{(L/L_0)^2},
\end{equation}
which works reasonably well with parameters $A_T = 4.82(26)$ and $L_0 = 8.515(89)$.
Increasing the rate of tolerated failures to, e.g., $p_f = 10\%$, merely reduces the
pre-factor to $A_T = 3.64(19)$, but leaves $L_0$ almost invariant. This data brings
back to attention the fact that, although the GEM approach works quite well, and
clearly outperforms simulated annealing, it naturally cannot evade the ${\cal
  NP}$-hard nature of the problem enforcing an exponential growth of computational
effort. To complicate the matter further, it is well conceivable that the shape
parameter $\xi$ of (\ref{eq:GEV}) increases with system size, as was observed in
tunneling simulations of spin-glass models \cite{alder:04}. Since for $\xi > 1/2$ the
variance of $G_{\xi,\mu,\sigma}$ becomes ill-defined, and for $\xi > 1$ additionally
the mean diverges, this would imply that a correct choice of population size ${\cal
  S} = T$ for all disorder configurations becomes impossible beyond a certain system
size.

\subsection{Hardness of samples and refinements\label{sec:hardness}}

Numerical ground-state computations in (non-polynomial) spin-glass systems are
subject to two types of ``hardness problems'': the exponential growth of {\em
  average\/} computational effort with system size and the large fluctuations in
sample hardness implied by heavy-tailed distributions of the type (\ref{eq:GEV}).
While ${\cal NP}$-hardness means exponential effort for the worst-case samples, it is
clear that (close to) ferromagnetic (i.e., best case) configurations can be tackled
in polynomial time. Hence, the difference in effort diverges with system size. While
this is true for the set of all {\em possible\/} input data, it is not clear {\em a
  priori\/} that the samples receiving non-negligible weight from the $P(\{J_{ij}\})$
considered indeed show such spread as implied by the data of Fig.~\ref{fig:GEV}. In
this Section, I discuss technical refinements of the GEM technique designed to
address the problem of large fluctuations in hardness.

\subsubsection{Effort adaptation}

\begin{figure}[t]
  \centering
  \includegraphics[keepaspectratio=true,scale=0.75,trim=45 48 75 78]{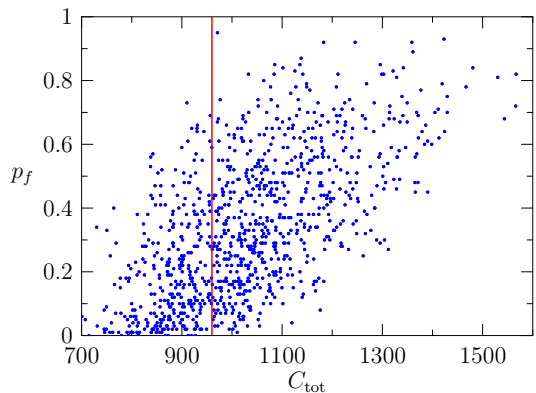}
  \caption
  {(Color online) Correlation diagram of failure probabilities $p_f$ for GEM runs of
    length ${\cal S} = 64$ and the total number of crossovers $C_\mathrm{tot}$ in
    ODGEM (overlap-driven GEM) runs with ${\cal D} = 0.128$ for $16\times 16$
    disorder samples. The vertical line shows the total number $C_\mathrm{tot} = 960$
    of crossovers for the GEM runs.
    \label{fig:success_runtime_correlation}}
\end{figure}

The fixed total number of crossovers $C_\mathrm{tot} = 2{\cal C}({\cal S}-4)$
performed by the GEM algorithm of Sec.~\ref{sec:genetic_matching} is not optimal in
view of the hardness variations observed. Additionally, one needs to tune ${\cal C}$
for best performance. It turns out that, in fact, the number of crossovers can be
determined automatically and on the run. This is done by comparing each pair of
configurations generated by crossover and potential replacement of the parents: if
they are too similar, one of them is removed from the population. The similarity is
here again measured by the optimized scalar overlap $\hat{q}^{\alpha\beta}$ resulting
from Eq.~(\ref{eq:overlap_matrix}), using a cutoff $q_\mathrm{max} = 1-2{\cal
  D}/L^2$. For an Ising spin glass, ${\cal D}$ would correspond to the number of
lattice sites where the two configurations disagree. For the algorithm of
Sec.~\ref{sec:genetic_matching}, this means that the loop over $c$ in lines 5 and 24
as well as the halving step in lines 25--32 are removed, whereas the following
instructions are inserted after line 23:

\vspace*{2ex}
\begin{algorithmic}[1]
  \setcounter{ALG@line}{23}
    \If{\Call{Overlap}{$\alpha, \beta$} $> 1-2{\cal D}/L^2$}
       \State remove configuration $\alpha$
       \State $s \gets s-1$
    \EndIf
\end{algorithmic}
\vspace*{2ex} This modified algorithm is referred to here as ODGEM (overlap-driven
GEM). This prescription ensures maximal genetic diversity at all times, and only
members able to produce ``novel'' configurations in crossover are retained in the
population.  As a consequence, the total number of crossovers is no longer fixed, but
depends on the used disorder configuration. It appears plausible that hard disorder
samples with many conflicting solutions close to the global minimum retain genetic
diversity longer than easy samples.  Figure~\ref{fig:success_runtime_correlation}
shows a correlation plot between the failure probabilities $p_f$ in the original GEM
and the total number of crossovers $C_\mathrm{tot}$ in the ODGEM approach (which is
proportional to the total computational effort). The Pearson correlation coefficient
\cite{feller:68},
\begin{equation}
  \label{eq:correlation_coefficient}
  \rho_{X,Y} = \frac{\langle [X-\langle X\rangle][Y-\langle Y\rangle]\rangle}
  {\sigma_X\,\sigma_Y}
\end{equation}
is found to be $\rho_{p_f,C_\mathrm{tot}} = 0.625(15)$, indicating clear but not
perfect correlation. As is seen from Fig.~\ref{fig:success_runtime_correlation}, most
samples with high failure rates in GEM receive an increased effort in ODGEM, and only
a few cases are missed. The choice of the overlap cutoff ${\cal D}$ must ensure that when
comparing configurations genetic ``equality'' is assumed only when no relative,
non-trivial excitation exists. Values of ${\cal D} = 0.1$ to ${\cal D} = 1$ are found
appropriate here.

\begin{figure}[t]
  \centering
  \includegraphics[keepaspectratio=true,scale=0.75,trim=45 48 75 78]{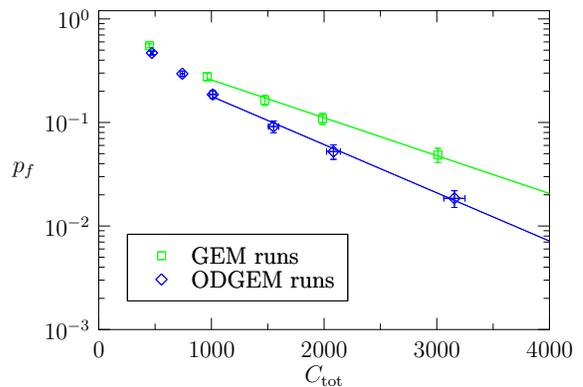}
  \caption
  {(Color online) Average failure probabilities for the GEM and ODGEM techniques on
    $16\times 16$ samples as a function of the total number $C_\mathrm{tot}$ of
    crossovers performed. The lines show fits of the functional form
    (\ref{eq:average_failure_law}) to the data.
    \label{fig:GEM_oGEM}}
\end{figure}

Since the ODGEM method detects a significant proportion of hard samples, the
distribution of the total number $C_\mathrm{tot}$ of crossovers acquires itself a
heavy tail, reflecting the hardness distribution described by Eq.~(\ref{eq:GEV}).
This empirical distribution is found to be well modeled by the extreme-value shape
(\ref{eq:GEV}) with $\xi$ close to zero, i.e., a Gumbel form. As is evident from the
data presented in Fig.~\ref{fig:GEM_oGEM}, this leads to an improved average
performance of the ODGEM compared to the GEM technique, increasing with the total
effort invested. Compared to GEM, ODGEM invests more effort in the hard samples and
less in the easy ones, as appears adequate. Direct inspection of the histogram
$H(p_f)$ of failure probabilities for the ODGEM method reveals that, indeed, the
number of instances with large failure probabilities are dramatically reduced as
compared to the GEM data of the same population size presented in
Fig.~\ref{fig:success_J}.

\subsubsection{Hardness indicators}

The possibility of deciding about sample hardness {\em a priori\/} could largely
increase the efficiency and reliability of the GEM (and any other ground-state
search) method. It is an open problem whether for spin-glass systems there are
microscopic sample properties significantly more easily computable than the ground
state itself which feature a strong correlation to sample hardness (with respect to a
given method) \cite{moore:87,kobe:95,mertens:02}. To investigate this question in the
context of the GEM approach and the 2D bimodal {\em XY\/} spin glass, I analyzed
correlations between the failure probabilities $p_f$ and a large number of
observables of the disorder sample, the candidate ground states found, and properties
of the population of configurations in the genetic algorithm. A number of resulting
correlation coefficients as estimated for $16\times 16$ samples from GEM and ODGEM
runs with population size ${\cal S}= 64$ are collected in
Tab.~\ref{tab:correlations}.

Regarding easily computable properties of the disorder realizations at hand, one
finds some moderately significant, positive correlation between $p_f$ in GEM runs and
the number of frustrated plaquettes, indicating increased hardness for highly
frustrated samples. Vice versa, an increasing average size of unfrustrated regions
leads to more and more success in finding ground states. Apparently, these
correlations are successfully taken into account by the modified algorithm ODGEM,
where no significant correlations between $p_f$ and sample properties are left. Also,
some properties of the computed (candidate) ground states correlate with success
probabilities.  In particular, a larger ground-state energy, indicating increased
frustration, is accompanied by a larger number of failures. Also, larger than average
values of the configurational r.m.s.\ chirality $\kappa$ \cite{kawamura:95},
\begin{equation}
  \label{eq:chirality}
  \kappa^2L^2 = \sum_{\square_n}\{\sum_{(i,j)\in\square_n} \sign J_{ij}\,
    [\bm{S}_i\times\bm{S}_j]_z\}^2,
\end{equation}
as well as the non-collinearity $Q$ \cite{saslow:86},
\begin{equation}
  \label{eq:noncollinearity}
  Q^4 L^4/2 = \sum_{\square_n}\sum_{(i,j)\in\square_n} |\bm{S}_i\times\bm{S}_j|^2,
\end{equation}
correlate with larger failure probabilities and hence indicate harder samples for the
method. Note that the definition(\ref{eq:chirality}) is specific to the case of
planar spins. For the Heisenberg model, the chirality is, instead, cubic in the spin
variables \cite{kawamura:92}. Again, such correlations are not (highly) significant
for the ODGEM technique, indicating that the corresponding samples automatically
receive higher computational effort there.

\begin{table}[t]
  \centering
  \renewcommand{\arraystretch}{1.1}
  \begin{tabular}{ldd}
    \toprule
    Observable & \rho_\mathrm{GEM} & \rho_\mathrm{ODGEM} \\
    \colrule
    No.\ AF bonds              &  0.020(36) &  0.019(34) \\
    No.\ frustrated plaquettes &  0.104(33) & -0.028(33) \\
    Size of unfrust. clusters  & -0.129(38) & -0.012(37) \\
    \colrule
    Energy                     &  0.226(33) &  0.061(34) \\
    Magnetization              &  0.021(34) &  0.024(36) \\
    Chirality                  &  0.137(34) &  0.001(33) \\
    Non-collinearity           &  0.266(33) &  0.101(33) \\
    \colrule
    Plain overlap              & -0.087(31) &  0.059(32) \\
    Optimized overlap          & -0.414(25) & -0.339(28) \\
    BED domain size            & -0.383(22) &  0.237(34) \\
    Parent replacements        &  0.328(29) &  0.297(34) \\ 
    \botrule
  \end{tabular}
  \caption
  {Estimated correlation coefficients between the failure probabilities $p_f$ in GEM
    and ODGEM runs and various observables of the disorder realizations, the actual
    ground-state configurations found, and the population of configurations within
    the genetic algorithm.
    \label{tab:correlations}}
\end{table}

The largest correlations with the failure probability are seen for properties of the
population of spin configurations inside of the GEM or ODGEM runs. The rather strong
inverse correlation between the average optimized overlap
$\langle\hat{q}^{\alpha\beta}\rangle$ of configurations and $p_f$ in GEM runs shows
that a homogeneous (i.e., large overlap) population occurs for a clear-cut, more
easily accessible ground state. Such homogeneity also implies that larger domains are
being identified in the BED decomposition. On the contrary, a large number of
successful replacements of parents by better offspring indicates stronger competition
of candidate ground states, resulting in more failures. These correlations related to
population remain, although weakened, in the maximum-diversity version ODGEM.
Consequently, devoting additional effort to disorder configurations singled out by
these population characteristics in ODGEM runs will additionally reduce failure rates
for hard samples, as I now discuss.

\subsubsection{Repeated runs}

Allocating such additional effort to allegedly hard samples typically means
performing additional, statistically independent runs to finally pick the
lowest-energy state found as the final result. As discussed above in
Sec.~\ref{sec:probabilistics}, however, the decrease in failure probability expected
from such a combination of several runs as described by Eq.~(\ref{eq:failure_law}) is
identical to the effect of performing a single computation with a larger population
(and this stays true for the modified ODGEM technique, at least to a good
approximation). In contrast to single, more expensive runs for all disorder
realizations, however, repeated runs allow to treat individual realizations
differently, in accord with the heavy-tailed distribution of sample hardness observed
in Fig.~\ref{fig:GEV}.

Even disregarding the use of the hardness indicators discussed in the previous
Section, however, it turns out to be beneficial to replace runs of length $T$ by a
number $n$ of shorter runs of length $T/n$: within the range of validity of
Eq.~(\ref{eq:failure_law}), the total probability $p_f$ not to find the ground state
remains unchanged. For an easy sample with small $p_f$, all but a small fraction of
the $n$ runs will end with a state of the same (namely the ground state) energy. For
hard samples with larger $p_f$, however, states with different energies will result
from a sizable fraction of runs, even if none of them is a ground state. In other
words, the structure of low-lying excited states typically results in the appearance
of a variety of energies for samples where a ground state is not found. By reacting
to these events with additional runs for the affected samples until the same minimum
energy has been found a certain number $n_0$ of times, the average failure probability
$\langle p_f\rangle$ can be further decreased.

To demonstrate the power of this extension, $n = 3$ runs of length $T = {\cal S} =
32$ for $1\,000$ samples of size $16\times 16$ were performed. For $722$ samples, all
three runs ended with the same minimum energy, which was consequently accepted as
estimate for the ground-state energy. For the remaining replica, an additional run
with ${\cal S} = 32$ was performed, which settled $74$ of the ``questionable'' cases.
This scheme was iterated until for all $1\,000$ samples the state of lowest energy
had been found $n_0 = 3$ times. The average effort for this computation corresponded
to ${\cal S} \approx 115$ (i.e., $3.6$ runs of length ${\cal S} = 32$), but the total
number of missed ground states corresponded to that of runs with ${\cal S} \approx
240$ (or $7.5$ runs of length ${\cal S} = 32$). This type of computation can, of
course, be favorably combined with the correlation results of
Tab.~\ref{tab:correlations} to perform a certain number of additional runs for
samples where certain observable values indicate especially large failure probability
$p_f$. Since the improvement effected by this addition depends on the structure of
low-lying excited states of the model, it unfortunately cannot be quantified in
general.

\section{Conclusions\label{sec:concl}}

I have presented a novel optimization heuristic for finding numerically exact ground
states of two-dimensional spin-glass systems with continuous spins with high
reliability. Embedding Ising spins into the continuous rotators, this exponentially
hard optimization problem is being related to the polynomial problem of finding Ising
ground states on planar graphs via Edmonds' algorithm \cite{edmonds:65a} for solving
minimum-weight perfect matching problems. Due to a history dependence of effective
coupling constants, this technique exhibits metastability on a low-energy subset of
the metastable states of a single spin-flip zero-temperature quench. In contrast to
simulated annealing and similar techniques, however, embedded matching has the
crucial advantage of being strictly downhill in energy. To find true ground states,
embedded matching is inserted as minimization procedure in a genetic algorithm
specially tailored to the spin-glass ground-state problem. The essential component
is here given by a properly chosen crossover operation exchanging automatically
determined domains of rigidly locked spins between the parent replica, thus
preserving the good optimization achieved at intermediate stages inside of domains
and effectively allowing the method to directly operate on the manifold of metastable
states.

This combination of techniques resulting in the genetic embedded matching (GEM)
method outperforms general-purpose approaches such as simulated annealing by orders
of magnitude: a success probability $p_s \ge 1\%$ could not be achieved at all with
reasonable computational effort with simulated annealing runs for the $16\times 16$
bimodal {\em XY\/} samples considered for performance comparison. Due to the
generally strong corrections to scaling present in spin-glass systems, the extension
in accessible system sizes effected by the GEM approach over general-purpose
techniques turns out to be crucial for the understanding of the asymptotic behavior
of the spin-glass phase, cf.\ Refs.~\cite{weigel:05f,weigel:06c}. The distribution of
success probabilities of the GEM technique can be understood from the decomposition
theorem (\ref{eq:failure_law}) of failure probabilities. The thus estimated
distribution over disorder replica of minimum required runtimes or population sizes
is perfectly described by a Fr\'echet distribution known from extreme-value theory,
which is plausible given that sample hardness is determined by the hardest of a
number of effective two-level systems describing the energy landscape of a disorder
realization.  Due to the heavy tail of this distribution, the exponential divergence
of average computational effort with system size expected from non-polynomial
optimization problems is accompanied by an increasing spread in sample hardness
impeding an appropriate choice of optimization parameters common to all disorder
samples. The variant approach ODGEM automatically maximizing genetic diversity by
monitoring configurational overlap, reduces the severity of this spread by devoting
additional effort to hard samples. Hard samples can also be detected by indicator
observables of the disorder and low-energy configurations as well as the population
in the genetic algorithm in order to decrease the failure probability in these cases.
The decomposition property (\ref{eq:failure_law}) finally allows to break up
computations in smaller units which, besides allowing to further reduce the average
effort required for a given success probability, makes the method ideally suitable
for computations on parallel workstation and Beowulf clusters.

Note that for systems with degenerate ground states, the GEM and ODGEM methods as
non-equilibrium techniques do not yield these different states with probabilities
proportional to the corresponding Boltzmann factors, such that in these cases an
additional post-processing of the states found would become necessary
\cite{hartmann:00a}. Recent evidence suggests, however, that such degeneracies might
be very unusual in disordered systems with continuous spins
\cite{weigel:05f,weigel:06c}. The performance analysis presented here focused on the
bimodal {\em XY\/} spin glass on the square lattice. The method straightforwardly
generalizes to any other nearest-neighbor O($n$) spin model (\ref{eq:EA_model}) on
planar graphs and to arbitrary disorder distributions. Specifically, the case of
Gaussian bond distribution can be treated with similar efficiency. Due to the use of
embedded matching as minimization component, the present from of the technique is
limited to planar, two-dimensional lattices and the case of zero field.  Other
minimization techniques might be used in lieu of embedded matching inside of the
genetic algorithm to tackle spin glasses in three dimensions or including magnetic
fields.

\begin{acknowledgments}
  I am indebted to M.\ Gingras for useful discussions and a critical reading of the
  manuscript. The research at the University of Waterloo was undertaken, in part,
  thanks to funding from the Canada Research Chairs Program (Michel Gingras). Some of
  the computations were performed at the facilities of the Shared Hierarchical
  Academic Research Computing Network (SHARCNET:www.sharcnet.ca). The author
  acknowledges support by the EC in form of a ``Marie Curie Intra-European
  Fellowship'' under contract No.\ MEIF-CT-2004-501422.
\end{acknowledgments}


\end{document}